\documentclass[12pt,preprint]{aastex}
\usepackage{times,pslatex}
\usepackage{amsmath,amsfonts}
\usepackage{graphicx}
\usepackage{epsfig}

\def\kms{\ifmmode{\rm km\thinspace s^{-1}}\else km\thinspace s$^{-1}$\fi}

\newcommand{\kepler}{{\it Kepler}~}
\newcommand{\keplerp}{{\it Kepler}}

\newcommand{\Mjup}{\mbox{$M_{\rm Jup}$}}

\newcommand{\Mplan}{\mbox{$M_{\rm p}$}}

\newcommand{\Pcbp}{\mbox{$P_{\rm CBP}$}}
\newcommand{\ap}{\mbox{${\rm a_{CBP}}$}~}
\newcommand{\aPCE}{\mbox{${\rm a_{CBP, PCE}}$}~}
\newcommand{\ep}{\mbox{${\rm e_{CBP}}$}~}
\newcommand{\ePCE}{\mbox{${\rm e_{CBP, PCE}}$}~}

\newcommand{\Msun}{\mbox{${\rm M_\odot}$}}
\newcommand{\Rsun}{\mbox{${\rm R_\odot}$}}
\newcommand{\abin}{\mbox{${\rm a_{bin}}$}~}
\newcommand{\ebin}{\mbox{${\rm e_{bin}}$}~}
\newcommand{\aCE}{\mbox{${\rm \alpha_{CE}}$}~}
\newcommand{\tCE}{\mbox{${\rm T_{CE}}$}~}
\newcommand{\Mprim}{\mbox{${\rm M_1}$}}
\newcommand{\Msec}{\mbox{${\rm M_2}$}}

\newcommand{\ac}{\mbox{$\alpha_{\rm crit}$}~}
\newcommand{\aM}{\mbox{${\rm \alpha_M}$}~}
\newcommand{\aalpha}{\mbox{$\alpha_{\rm 1}$}}
\newcommand{\aalphao}{\mbox{$\alpha_{\rm 0.1}$}}

\slugcomment{Accepted for publication in the Astrophysical Journal}
\shorttitle{Not for re-distribution.}

\begin{document}

\title{Tatooine's Future: The Eccentric Response of {\it Kepler's} Circumbinary Planets to Common-Envelope Evolution of their Host Stars}
\shorttitle{Dynamical Evolution of Circumbinary Planet}

\author{
Veselin~B.~Kostov\altaffilmark{1,6},
Keavin~Moore\altaffilmark{2},
Daniel~Tamayo\altaffilmark{3,4,5},
Ray Jayawardhana\altaffilmark{2},
Stephen~A.~Rinehart\altaffilmark{1}
}

\email{veselin.b.kostov@nasa.gov}

\altaffiltext{1}{NASA Goddard Space Flight Center, Mail Code 665, Greenbelt, MD, 20771}
\altaffiltext{2}{Faculty of Science, York University, 4700 Keele Street, Toronto, ON M3J1P3, Canada}
\altaffiltext{3}{Department of Physical \& Environmental Sciences, University of Toronto at Scarborough, Toronto, Ontario M1C 1A4, Canada}
\altaffiltext{4}{Canadian Institute for Theoretical Astrophysics, 60 St. George St, University of Toronto, Toronto, Ontario M5S 3H8, Canada}
\altaffiltext{5}{CPS Postdoctoral Fellow}
\altaffiltext{6}{NASA Postdoctoral Fellow}

\begin{abstract}

Inspired by the recent \kepler discoveries of circumbinary planets orbiting nine close binary stars, we explore the fate of the former as the latter evolve off the main sequence. We combine binary star evolution models with dynamical simulations to study the orbital evolution of these planets as their hosts undergo common-envelope stages, losing in the process a tremendous amount of mass on dynamical timescales. Five of the systems experience at least one Roche-lobe overflow and common-envelope stages (Kepler-1647 experiences three), and the binary stars either shrink to very short orbits or coalesce; two systems trigger a double-degenerate supernova explosion. {\it Kepler's} circumbinary planets predominantly remain gravitationally bound at the end of the common-envelope phase, migrate to larger orbits, and may gain significant eccentricity; their orbital expansion can be more than an order of magnitude and can occur over the course of a single planetary orbit. The orbits these planets can reach are qualitatively consistent with those of the currently known post-common-envelope, eclipse-time variations circumbinary candidates. Our results also show that circumbinary planets can experience both modes of orbital expansion (adiabatic and non-adiabatic) if their host binaries undergo more than one common-envelope stage; multiplanet circumbinary systems like Kepler-47 can experience both modes during the same common-envelope stage. Additionally, unlike Mercury orbiting the Sun, a circumbinary planet with the same semi-major axis can survive the common envelope evolution of a close binary star with a total mass of $1~\Msun$. 

\end{abstract}


\section{Introduction}

{\it ``Everything is in motion''}, Plato quotes Heraclitus (Cratylus, Paragraph 402, section a, line 8), {\it ``and nothing remains still.''}. Despite the overwhelming timescales, from a human perspective, the natural life-cycle of stars is a prime example of change on the cosmic stage. Following the laws of stellar astrophysics, over millions to billions of years, stars and stellar systems form, evolve, and ultimately die (e.g. \citealt{kipp90} and references therein). And so do planetary systems, with their fate linked intimately to that of their stellar hosts. 

The fate of planets orbiting single stars has been studied extensively, indicating that planets can survive their star's evolution if they avoid engulfment and/or evaporation during the Red Giant Branch (RGB) and the Asymptotic Giant Branch (AGB) stages (e.g.~\citealt{livio84,rasio96,duncan98,villaver07}). There is also accumulating evidence of planetary or asteroidal debris surrounding white dwarfs and polluting their atmospheres (see recent reviews by \citealt{farihi16} and \citealt{veras16}). A planet's survival depends both on the initial orbital separation, and on its mass. For example, even a planet as massive as 15 \Mjup~ around an 1\Msun~MS star can be destroyed inside the stellar envelope during the AGB phase \citep{villaver07}. An unpleasant prospect for the future of our own planet is that it might not survive the evolving Sun (e.g.~\citealt{schroder08}). Despite the complications, theoretical considerations indicate that planets can indeed survive the evolution of single stars \citep{villaver07}, and there is mounting observational evidence supporting this (e.g.~\citealt{reffert15,wittenmyer16}, and references therein\footnote{Also see https://www.lsw.uni-heidelberg.de/users/sreffert/giantplanets.html for a list of known systems.}).

Single stars, however, do not have monopoly over planetary systems. Nearly half of Solar-type stars are members of binary and higher order stellar systems \citep{raghavan10}, and an increasing number of planets have been discovered in such systems. While the presence of a distant stellar companion will have little effect on a planet around one (evolving) member of a wide binary stellar system ($\gtrsim100$~AU), a planet orbiting around both members of a close binary system (separation $\sim10$~AU or less) will experience qualitatively and quantitatively different stellar evolution. Namely, where a single star loses mass and expands on timescales of millions to billions of years \citep{veras16} while on the main sequence (MS), the RGB and the AGB -- its planets react accordingly -- close binary stars can experience events of dramatic mass loss, orbital shrinkage on timescales of years, and common envelope (CE hereafter, see Appendix for abbreviations and parameters) stages where the two stars share (and quickly expel) a common atmosphere \citep{paczynski76,hilditch01}. During the in-spiralling CE phase, the two stars can strongly interact with each other by transferring mass, smoothly coalesce or violently collide, or even explodes as a supernova (SN). Both the amount of energy released during this stage and the timescale of the release are staggering -- a close binary star can lose an entire solar mass over the course of just a few months (e.g.~\citealt{rasio96, livio84, ivanova13, passy12, ricker08, ricker12, nandez14}, but also see \citealt{sandquist98} and \citealt{demarco03} for longer timescales); in the extreme case of a SN, the binary will be completely disrupted. Overall, the evolution of close binary stars is much richer compared to single stars, as the separation, eccentricity and metallicity of the binary add additional complications to an already complex astrophysical process. Interestingly, to date there are 9 close main-sequence binary systems harboring confirmed circumbinary planets (hereafter CBPs, \citealt{doyle11, welsh12, welsh15, orosz12a, orosz12b, kostov13, kostov14, schwamb13}. It is reasonable to assume that the reaction of these planets to the violent CE phase of their host binary stars will be no less dramatic. 

Previous studies of the dynamical response of planets in evolving multiple stellar systems have focused on post-CE (hereafter PCE) CBP candidates (e.g.~\citealt{volschow13, portegies13, mustill13}), of circumprimary planets in evolving wide binary systems \citep{kratter12}, and of the survival prospects of CBP planets \citep{veras11, veras12}. In particular, \cite{portegies13} and \cite{mustill13} start with the final outcome of close binary evolution -- PCE systems with candidate CBP companions (HU Aqu and NN Ser respectively) -- and reconstruct the initial configuration of their progenitors. 

Inspired by recent discoveries of planets in multiple stellar systems, here we tackle the reverse problem -- we start with the 9 confirmed CBP \kepler systems with known initial binary configurations (using data from NASA's \kepler mission), and study their dynamical response as their stellar hosts evolve. We combine established, publicly-available binary stellar evolution code (BSE; \citealt{bse02}) with direct N-body integrations (using REBOUND, \citealt{rein12}) allowing for stellar mass loss and orbital shrinkage on dynamical timescales. For simplicity, here we focus explicitly on dynamical interactions only (e.g.~no stellar tides), and assume that the CBPs do not interact with the ejected stellar material. We expand on the latter limitation in Section \ref{sec:limitations}. 

We note that \cite{mustill13} assume adiabatic mass-loss regime \citep{hadjidemetriou63} during the CE stage for the evolution of NN Ser. In this case, the orbital period of the CBP is much shorter than the CE mass-loss timescale (hereafter ${\rm T_{CE}}$), and the planet's orbit expands at constant eccentricity\footnote{In line with thermodynamics nomenclature, this is an isoeccentric process.}. \cite{volschow13} study the two analytic extremes of mass-loss events for the case of NN Ser -- the adiabatic regime, and the instantaneous regime where \tCE is much shorter than the orbital period of the planets\footnote{We note that both the total amount of mass lost, and the rate at which it is lost will affect the evolution of a CBP's orbit during a CE phase.}. The \kepler CBP systems, however, occupy a unique parameter space where the planets' periods are comparable to the \tCE used by \cite{portegies13} and suggested by \cite{ricker08, ricker12} for binary systems similar to \kepler CBP hosts. Thus the adiabatic approximation may not be an adequate assumption when treating the CE phases of these systems. Specifically, the CE phase can occur on timescales as short as one year \citep{paczynski76, passy12}, and during this year a binary can expel $\sim2\Msun$ worth of mass \citep{ricker08, ricker12, iaconi16}. Interestingly, such timescales are well within a single orbit of CBPs like Kepler-1647 with \Pcbp = 3 yrs \citep{kostov16}. Such a dramatic restructuring will have a profound impact on the dynamics of the system -- for example, a CBP can achieve very high eccentricity -- and may even result in the ejection of the planet. Similar to \cite{portegies13}, here we tackle such complications by directly integrating the equations of motions during the CE phases for each \kepler CBP system. 

The transition from adiabatic to non-adiabatic regime can be characterized through a mass-loss index $\psi$. For a planet orbiting a single star \cite{veras12} define: 

\begin{equation}
\label{eq:PSI}
\Psi\equiv\frac{\aM}{n\mu} = (2\pi)^{-1}\left(\frac{\alpha_{\rm M}}{1\Msun~{\rm yr^{-1}}}\right)\left(\frac{\rm a_{p,0}}{1~{\rm AU}}\right)^{\frac{3}{2}}\left(\frac{\mu}{1~\Msun}\right)^{-\frac{3}{2}}
\end{equation}

{\noindent where ${\aM}$ is the mass-loss rate, ${\rm a_{p,0}}$ is the initial semi-major axis of the planet, and ${\rm \mu= M_{\rm star} + M_{\rm p}}$ is the total mass of the system\footnote{Throughout this paper we use subscripts ``bin'',  and ``p'' or ``CBP'' to denote the binary star and the CBP, and ``${\rm M_1}$'' and ``${\rm M_2}$'' to denote the mass of the primary and the secondary star respectively.}. If $\Psi \ll1$ then the evolution of the planet's orbit is in the adiabatic regime and its semi-major axis grows at constant eccentricity. 

Alternatively, if $\Psi\gg1$ then the planet sees an `instantaneous' mass loss and its orbital evolution is in a `runaway' regime. In this scenario, the fate of the planet depends on the ratio between the final and initial mass of the system, i.e. ${\rm \beta= \mu_{\rm final}/\mu_{\rm init}}$, and on the orbital phase of the planet at the onset of the mass-loss event. For example, if a highly-eccentric planet is at pericenter at the beginning of the CE stage it becomes unbound. Overall, ejection occurs when the system loses sufficient mass by the end of the mass loss event. To first order, the critical mass ratio for ejection ${\beta_{\rm eject}}$ is given by Eqn.~39 of \cite{veras11}, which we reproduce here for completeness:

\begin{equation}
\label{eq:ejection}
\beta_{\rm eject}\equiv{\rm 0.5(1+e_{p,0})}
\end{equation}

{\noindent where ${\rm e_{p,0}}$ is the eccentricity of the planet at the beginning of the mass-loss event. A planet on a circular orbit or at pericenter is ejected from the system if $\beta<\beta_{\rm eject}$\footnote{Also see Eqn.~48 from Veras et al.~(2011) for a comprehensive treatment.}. A planet at apocenter remains bound even in the runaway regime, its orbit expands to ${\rm a_{runaway, apocenter}}$ and circularizes, or reaches an eccentricity of ${\rm e_{(post-circ)}}$. If a planet remains bound in the runaway regime, it's semi-major axis is given by Eqn.~43 and 44 of \cite{veras11}:

\begin{equation}
\label{eq:ejection}
{\rm a_{runaway}\equiv\frac{\rm a_{p,0}(1 \mp e_{p,0})}{2 - \beta(1\pm e_{p,0})}}
\end{equation}

{\noindent where the signs represent initial pericenter/apocenter respectively.}

According to the above prescription, the dynamical evolution of a CBP's orbit for $\Psi \sim0.1-1$ lies between these two regimes. For both $\Psi \sim 1$ and $\Psi \gg 1$, the evolution is not adiabatic and \ep can vary (increase or decrease) as \ap varies. Thus the planet may become unbound only if $\Psi \gtrsim 1$, with the caveat that the transition is not clear-cut and the $\Psi\sim0.1-1$ regime is more complex \citep{portegies13, veras11, veras12, veras16}. As we show below, the fact that the central object in a CBP system is itself a binary instead of a single star adds yet another layer of complication to these theoretical considerations.

Based on Eqn. \ref{eq:PSI}, and assuming CE timescale of $\tCE\sim10^3$ yrs, we would a priori expect that all \kepler CBPs remain bound to their hosts during their primary stars' CE stage (e.g.,~see Fig. 3, \citealt{veras12}). However, for the rapid CE timescale of $\tCE\sim1$~yr suggested by recent results (e.g.~\citealt{ricker08, ricker12}), some of the \kepler CBP systems are within a factor of 5-10 of the transition to non-adiabatic regime during the primary or secondary CE (i.e. $\Psi\sim0.1-0.2$), and within a factor of 2 of the transition during their respective secondary CE stage (i.e. $\Psi\sim0.5$). We might therefore expect significant variations in the respective CBPs' orbital eccentricities, as well as deviations from the final semi-major axes expected from adiabatic mass loss. 

As the transition between adiabatic and non-adiabatic regimes is difficult to study analytically, here we examine the dynamical evolution of all \kepler CBP systems numerically, using the adaptive-timestep, high-order integrator IAS15 \citep{rein15}, which is available as part of the modular and open-source REBOUND package \citep{rein12}. REBOUND is written in C99 and comes with an extensive Python interface.

This paper is organized as follows. In Section \ref{sec:bse} we describe the algorithm we use for the evolution of the \kepler CBP hosts. Section \ref{ref:rebound} details our implementation of an N-body code in exploring the dynamical reaction of said CBPs to the respective CE stages. We present the results for each system in Section \ref{sec:results}, and discuss our results and draw conclusions in Section \ref{sec:end}. 

\section{Binary Star Evolution}
\label{sec:bse}

The evolution of a gravitationally-bound system of two stars depends primarily on its initial configuration. If the binary is wide enough the stars will evolve in isolation, according to the prescription of single-star evolution theory. Alternatively, if the initial separation is sufficiently small then the stars will influence each other's evolution. The evolution of such systems, described as close (or interacting) binary stars \citep{hilditch01}, is more complex and includes a number of additional processes such as tidal interactions, mass transfer, etc. The 9 \kepler CBP-harboring binary systems fall in the latter category, and we use the established, open-source Binary Star Evolution code \citep{bse02} to study their evolution. Briefly, the code works as follows.

While the binary is in a detached state, the code evolves each star individually using a single star evolutionary code \citep{sse00} which includes tidal and braking mechanisms, and wind accretion. When the stars begin to interact, BSE uses a suite of binary-specific features such as mass transfer and accretion, common-envelope evolution, collisions and mergers, and angular momentum loss mechanisms. The evolution algorithm allows the specific processes to be turned on and off, or even modified based on custom requirements. By default, all stars are assumed to be initially on the zero-age main sequence (ZAMS), but any possible evolutionary state, such as corotation with the orbit, can be the starting point of the simulation. BSE uses an evolution time-step small enough to prevent the stellar mass and radius from changing too much (not more than $1\%$ in mass and $10\%$ in radius), which allows identifying the time when, and if, a star first fills its Roche lobe.  

At the onset of Roche lobe overflow (RLOF), the two stars either come into contact and coalesce, or initiate a common envelope (CE) stage. CE evolution is a complex mechanism that typically occurs when an evolved, giant star transfers mass to a main sequence (MS hereafter) star on a dynamical timescale\footnote{A collision between a dense core and a star can also trigger a CE phase.}. When the giant star overfills the Roche lobe of both stars, its core and the MS star share a single envelope. This envelope rotates slower than the orbiting ``cores'' within due to expansion, and the resulting friction causes in-spiral and energy transfer to the envelope, described by an efficiency parameter \aCE. The outcome of this phase is either ejection of the envelope (assumed to be isotropic) if neither core fills its Roche lobe, leaving a close white dwarf-MS pair (assumed to be in corotation with the orbit), or coalescence of the cores. A challenging process to study numerically, BSE recognizes the CE phase -- beginning with RLOF -- and forces the system through it on an instantaneous timescale, allowing the evolution to continue according to the appropriate PCE parameters. The code resolves the outcome of the CE phase based on the initial binding energy of the envelope and on the initial orbital energy of the two cores\footnote{BSE also accounts for collisions which do not proceed through CE evolution, and for the possibility of core-sinking depending on the stellar types (which may lead to rejuvenation).}.

On input BSE requires a number of parameters. Namely, the CE efficiency parameter \aCE (in the range of $0.5-10$), the masses and stellar types of both the primary and secondary stars (in the range of $0.1 - 100 \Msun$), the binary orbital period (in days), eccentricity ($0.0 - 1.0$), and the maximum time for the evolution (here we choose 15 Gyr, roughly the age of the Universe, unless a particular \kepler CBP system stops evolving -- or the stars coalesce -- much sooner, at which point we stop the simulations), and the metallicity ($0.0001 - 0.03$) of the system. The code also checks whether tidal evolution is included, i.e. ``on'' (hereafter TCP for tidal circularization path) or ``off'' (NTCP for no tidal circularization path). Given the still-uncertain mechanism of the CE stage, the parameters describing the efficiency of envelope ejection and the treatment of tidal decay -- and their associated uncertainties -- are likely the dominant source of error in our BSE results. 

Depending on \aCE, a binary system may exit a CE stage as a merged star, as a very tight PCE binary star, or may even trigger a Supernova explosion. For example, a higher \aCE accounts for energy sources other than orbital energy, and the system then requires less energy to dissipate the envelope. The BSE code also allows an alternate CE model to be used, the de Kool CE evolution model \citep{kool90}, which first introduced the CE evolution binding energy factor $\lambda$.

Likewise, tides are integral to the evolution of close binary stars. The strength of tidal dissipation will affect the orbital separation and eccentricity when mass transfer begins, and hence the future evolution under RLOF or a CE stage. BSE achieves orbital circularization mainly through tides, but also by accounting for more mass accretion at pericenter than apocenter of an eccentric orbit. Another effect is the exchange of orbital angular momentum with the component spins due to tidal synchronization, thus causing the evolution to proceed in a similar way to closer binary systems if tides were ignored. In addition, primary spin-orbit corotation can be forced within the code to avoid unstable RLOF, which gives two possible outcomes for each system. To account for the effect of tides on binary evolution, we use both the tides ``on'' (i.e. TCP) or ``off'' (i.e. NTCP) options in BSE. As a result, the binary starts the first RLOF at either  ${\rm e_{bin,RLOF} = 0}$ and ${\rm a_{bin,RLOF} < a_{\rm bin, 0}}$ (for TCP), or at ${\rm e_{bin,RLOF} = e_{\rm bin, 0}}$ and ${\rm a_{bin,RLOF} = {\rm a_{bin, 0}}}$ (for NTCP).

Additionally, while the binary periods of the \kepler CBP systems are known to very high precision, their stellar masses and metallicities have non-negligible uncertainties\footnote{Interestingly, their metallicities are typically sub-Solar.}. For example, the primary masses for Kepler-38 and -64 have $1\sigma$ errors of ${\rm 0.05~M_{Sun}~and~0.1~M_{Sun}}$ respectively, and the ${\rm 1\sigma}$ errors on the metallicities of half of the systems are as large as, or even larger than the measured values. To better characterize the impact of these uncertainties on the evolution of the \kepler CBP systems, we obtain BSE results for both the best-fit mass and metallicity values and also for their corresponding $3\sigma$ range\footnote{The BSE upper limit of $Z = 0.03$ restricts some systems with greater $Z$, such as Kepler-64, to this extreme.}. 

The entire parameter space we explored, shown in Table \ref{tab:bseparams}, includes a range of CE parameters as well and consists of 22680 BSE simulations. Our default configurations are denoted in boldface in the table; for the rest of the input parameters we use the default values from model A of \cite{bse02}, described as the most favorable and effective. Model A uses all default wind-related parameters, and a Reimers mass-loss coefficient of $\eta = 0.5$. 

As BSE evolves the stars, their types change, and each type is checked throughout the code to ensure the correct evolutionary tracks are used for both the individual stars as well as the system as a whole. These numerical values, as well as what they represent, are listed in Table \ref{tab:bse_stellar_types}. At each significant event, e.g. at the beginning and end of the RLOF, and of the CE phase, when the stars coalesce or collide, the code also produces a ``type'' label. These labels, along with their meaning, are shown in Table \ref{tab:bseoes}.

An example of the BSE output for Kepler-38, \aCE = 0.5, NTCP and TCP is shown in Table \ref{tab:BSE_example}. The BSE results for all \kepler systems can be found in an online supplement.   

In the following section we describe how we use the initial and BSE-generated final binary masses and orbital separations for each {\it Kepler} CBP system to explore the dynamical response of the planets during the respective CE stages.

\section{Dynamical Simulations}
\label{ref:rebound}

While the concept of CE evolution for binary stars has been proposed 40 years ago \citep{paczynski76}, detailed understanding of this important stage is a challenging task that requires computationally-intensive hydrodynamical calculations. The outcome of this complex phase has been studied both numerically and analytically (see Table 1 of Iaconi et al. 2016 for a list of references). For example, hydrodynamical simulations by \cite{ricker12} for a 1.05\Msun+0.36\Msun~binary star with an initial orbital period of 44-days (similar to \kepler CBP hosts) show that in the later stages of the CE the system losses mass at a rate of 2~${\rm \Msun ~yr^{-1}}$, the binary orbital separation decreases by a factor of 7 after $\sim56$ days, and $\sim90\%$ of the outflow is contained within 30 degrees of the binary's orbital plane. 

To numerically account for the dynamical impact on \kepler CBPs from such drastic orbital reconfigurations and tremendous, short-timescale mass-loss phases, we use REBOUNDx -- a library for incorporating additional effects beyond point-mass gravity in REBOUND simulations ({\url{https://github.com/dtamayo/reboundx}). In particular, we shrink the binary orbits with the REBOUNDx {\tt modify\_orbits\_forces} routine which adds to the equations of motion a simple drag force opposite particles' velocity vectors \citep{papaloizou00}. When orbit-averaged over a Keplerian orbit, this approach results in an exponential damping of the semimajor axis where the e-folding timescale is set through the particle parameter ${\rm \tau_a}$. Hydrodynamical simulations by \cite{passy12} and \cite{iaconi16} exhibit similar orbital decay during the CE stage (e.g.~see their respective Figures 5 and 14), validating our approach. 

Given the complexities and uncertainties of the mass-loss mechanism during the CE phases, we adopt a simple analytical parametrization similar to the method of \cite{portegies13} who tackled the problem by using a constant mass-loss rate. Here we choose to instead use the same functional forms for the CE-driven changes in ${\rm M_{star}}$ as we do for \abin (utilizing the REBOUNDx {\tt modify\_mass} routine), i.e.~an exponential mass-loss for particles on their assigned e-folding timescales ${\rm \tau_M}$. Namely, for a given CE stage where one of the stars transitions from an initial mass $M_0$ to a final mass $M_f$ (both provided by BSE), the mass loss occurs over a time ${\tCE = (M_f - M_0)/{\aM}}$. We note that an exponential mass loss has been explored by \cite{adams13} and \cite{adams13b} as well, but for the case of single-star systems. 

To ensure a smooth mass evolution (and avoid possible numerical artifacts) that asymptotes at $M_f$, we linearly evolved REBOUNDx's exponential mass-loss rate so that it vanished at time \tCE.  In particular, in terms of the e-folding timescale $\tau_M$, 
\begin{eqnarray} \label{taum}
\dot{M} = - \frac{M}{\tau_{\rm M}},
\tau_{\rm M} = \frac{\tau_{\rm M,0}}{\rm (1 - t/\tCE)}
\end{eqnarray}
This differential equation can be solved analytically, yielding
\begin{equation}
M(t) = M_f\left( \frac{M_0}{M_f} \right)^{\rm (1 - t/\tCE)^2}
\end{equation}
as long as $\tau_{M,0}$ in Eq. 4 is chosen as
\begin{equation}
\tau_{\rm M,0} = \frac{\tCE}{2\ln(\frac{M_0}{M_f})},
\end{equation}

As the star's mass decreases from $M_0$ to $M_f$ during the respective CE stage, the binary orbit must shrink from ${\rm a_{bin,0}}$ to ${\rm a_{bin,PCE}}$ (as provided by BSE) as well -- and both must occur over a time \tCE. We apply the same prescription for evolving \abin as that for the mass-loss (i.e. Eqn. 4-6 with $\tau_{\rm a}$ instead of $\tau_{\rm M}$) but caution that while this would evolve \abin from ${\rm a_{bin, 0}}$ to ${\rm a_{bin,PCE}}$ in isolation, the mass-loss is simultaneously increasing \abin. To correct for this, and achieve the desired ${\rm a_{bin,PCE}}$ we manually adjust the corresponding $\tau_{\rm a,0}$ parameter. 

While tides are not directly accounted for in our dynamical simulations as the particles involved are treated as point masses, we capture their effects in a simple parametrized way. Specifically, we impose the same exponential decay to the evolution of ${\rm e_{bin}}$ as we do for the binary's semi-major axis shrinking and stellar mass-loss (i.e. Eqn. 4-6 but with $\tau_{\rm e}$). As a result ${\rm e_{bin}}$ undergoes dampened oscillations during the CE phase at the end of which it settles within a few percent of zero -- as required by BSE.

As mentioned in the previous section, while the BSE code yields initial and final values for the stellar masses and \abin at the beginning and end of the CE phases, it does not resolve the extremely rapid mass-loss phase. We bridge this gap by exploring two regimes of mass-loss, namely ${\rm \aalpha = 1.0~M_\odot / yr}$ and ${\rm \aalphao = 0.1~M_\odot / yr}$, denoted by their respective subscripts. The former is based on the results of \cite{ricker08, ricker12} and \cite{portegies13} for systems similar to the \kepler CBP hosts. Values of \aalphao~(and smaller) typically guarantee that the orbital evolution of these CBPs is in the adiabatic regime (see Eq. \ref{eq:PSI}). 

To test the applicability of our numerical simulations to the stated problem, we examine the evolution of a mock Kepler-38-like system where the binary is replaced with a single star of the same total initial and final mass according to the BSE prescription for Kepler-38 (see Table \ref{tab:BSE_example}), and using the same initial orbit of the planet. Here we test two mass-loss regimes -- runaway (for ${\aM = 100.0~\Msun/yr}$) and adiabatic (for ${\aM = 0.1~\Msun/yr\equiv\aalphao}$). The results are shown in Figure \ref{fig:K38_single_star}, where the three panels represent (from upper to lower respectively) the evolution of the stellar mass, of the planet's semi-major axis and of the planet's eccentricity as a function of time (such that mass loss starts at ${\rm  t = 0}$ and ends at ${\rm t = T_{mass-loss}}$). According to the analytic prescription, the planet should remain bound even in the runaway mass-loss regime as the mass lost is smaller than the critical mass limit for ejection -- indicated by the dotted line in the upper panel of the figure\footnote{For the planet to be ejected, the final stellar mass must be smaller than ${\rm M_1(runaway, eject)}$.}. The horizontal dashed lines in the middle and upper panels represent the theoretical adiabatic (green) and runaway (red) approximations. As seen from these two panels, the numerical simulations are fully consistent with the theoretical approximations, demonstrating the validity and applicability of our simulations. 

\begin{figure}
\centering
\epsscale{1.1}
\plotone{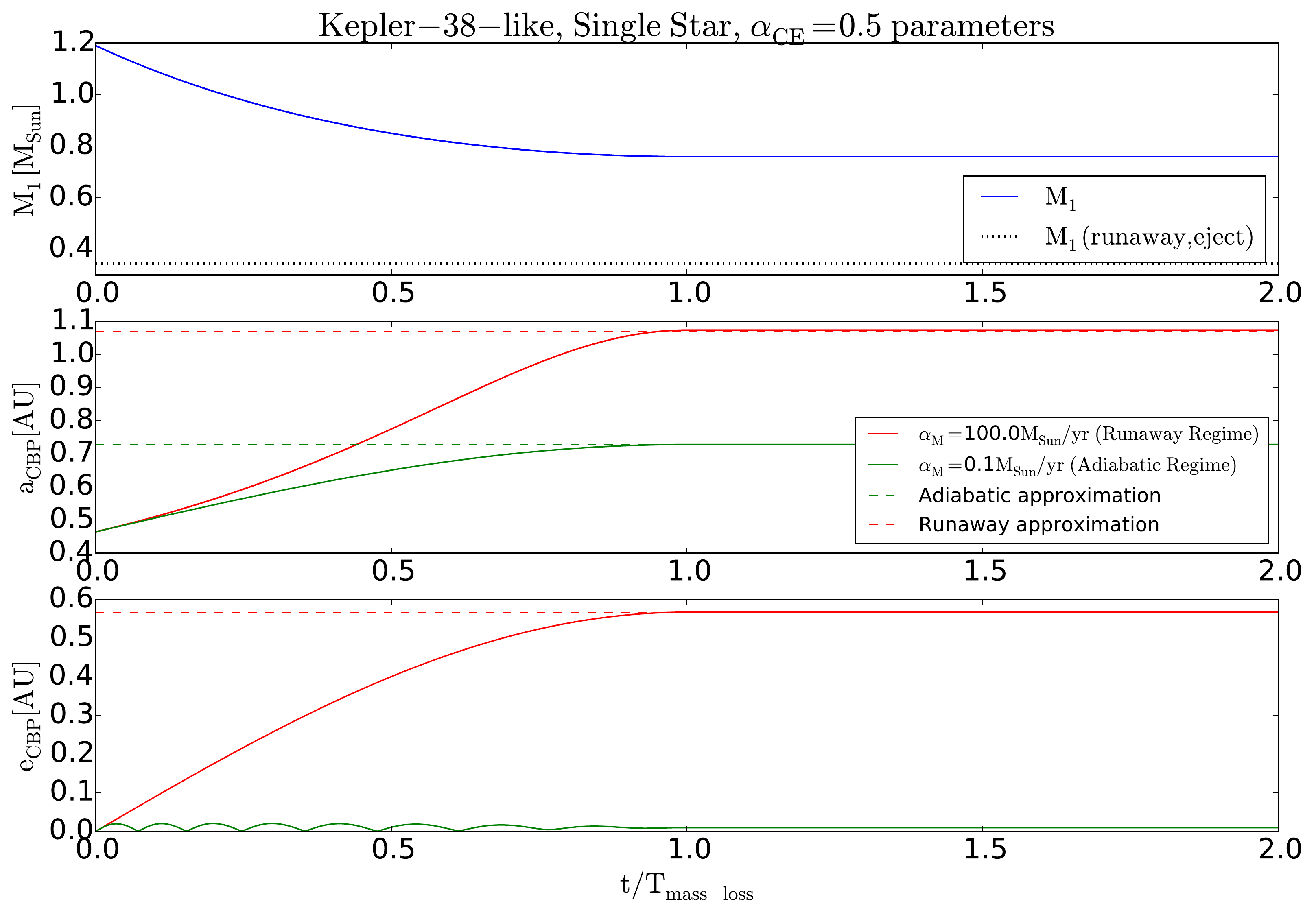}
\caption{Evolution of, from top to bottom, the stellar mass, the CBP semi-major axis, and eccentricity for a Kepler-38-like system, where the binary is replaced with a single star which loses mass according to the $\aCE = 0.5$ BSE simulation (for the primary CE phase) for two mass-loss regimes -- runaway (red) and adiabatic (green). Here ${\rm T_{mass-loss}}$ is the same as the corresponding \tCE for Kepler-38, \aCE = 0.5. The black dotted line in the upper panel represents the critical mass lost which would result in the ejection of the planet. The dashed lines in the middle and lower panels indicate the runaway (red) and adiabatic (green) approximations. The numerical evolution of the planet's semi-major axis and eccentricity is fully consistent with the theoretical expectation, validating the applicability of our simulations.
\label{fig:K38_single_star}}
\end{figure}

In order to interpret the results from numerical integrations, it is useful to define a critical mass loss rate, \ac, corresponding to $\Psi = 1$ (see Eq. \ref{eq:PSI}). CBPs in systems with ${\rm \alpha_M \gg \ac}$ will experience `instantaneous' mass loss, while the orbits of those in systems with ${\rm \alpha_M \ll \ac}$ should evolve adiabatically.  A CBP in a system where ${\rm \alpha_M}$ becomes comparable to or larger than \ac at any point during the binary evolution risks becoming unbound. Secondary CE stages will have two different \ac values as the respective CBP will have two different values for \ap and \ep (but the same $\mu$) at the end of the preceding CE stages (see Eq. \ref{eq:PSI}) -- one for \aalpha, and another for \aalphao. We denote these critical values as $\ac(\alpha_1)$ and $\ac(\alpha_{0.1})$. In the subsections below, for each CBP \kepler system we compare \ac to $\alpha_1$ and $\alpha_{0.1}$, for each CE phase. Where the adopted $\aalpha$ is within a factor of 10 of the respective \ac we pay special attention to the CBP orbit, and test its evolution for $0.1\aalpha$, $0.2\aalpha$, and $0.5\aalpha$ as well. Unless indicated otherwise, the final fate of the CBP for these cases is similar to the case for $\aalpha$.
 
The description of a three body system containing an evolving binary star is inherently multidimensional, and minor changes in one parameter can cascade into major changes in the rest. In addition to the importance of ${\rm \alpha_{CE}}$, tides and ${\rm \alpha_{M}}$, another key issue -- as shown in the next section -- is the phase offset between the binary star and the CBP at the onset of the CE phase, ${\rm \Delta\theta_0\equiv\theta_{CBP,0}-\theta_{bin,0}-\omega_{bin}}$ (i.e. the difference between their true anomalies, taking into account the binary's argument of the pericenter). 

Thus for each system we test four initial binary phases corresponding to the binary orbit turning points, i.e. Eastern and Western Elongations (EE and WE), Superior and Inferior Conjunctions (SC and IC) and 50 initial CBP phases (between 0 and 1, with a step of 0.02) for a total of 200 initial conditions. We explore each of these initial conditions for ${\rm \alpha_{CE} = (0.5, 1.0, 3.0, 5.0, 10.0)}$, for TCP/NTCP, and for \aalpha~and \aalphao.
 
Finally, we note that the post-CE semi-major axis of a CBP (${\rm a_{CBP,PCE}}$) that was initially on an eccentric orbit (i.e. ${\rm e_{CBP,0}} > 0$) depends on its orbital phase at the onset of the CE stage. For those \kepler systems that undergo a secondary CE phase with mass-loss, we start the respective integrations at the modes of ${\rm a_{CBP,PCE}}$ and ${\rm e_{CBP,PCE}}$ from the preceding CE stage.
 
\section{Results}
\label{sec:results}

Here we describe the results of our simulations for those \kepler CBP systems that experience at least one CE phase. Unless specified otherwise, the evolution of each system is for the default CE parameters listed in Table \ref{tab:bseparams}. The initial orbital parameters for each system are listed in Table \ref{tab:keplerinitial}. We note that our N-body integrations only cover the common envelope phases.

Below we outline the major binary evolution stages (as produced by BSE), along with the dynamical responses of the CBPs; the main results are presented in Tables \ref{tab:temp_tab_EB}, \ref{tab:temp_tab_CBP_fast}, \ref{tab:temp_tab_CBP_fast_cont}, \ref{tab:temp_tab_CBP_slow}, and \ref{tab:temp_tab_CBP_slow_cont}. All reported times are in Gyr since ZAMS. For simplicity, we assume that the current \kepler CBP systems are at ZAMS and that all CBP start on initially circular, coplanar orbits. 

Overall, five of the nine \kepler CBP systems experience at least one CE phase for their default metallicities (see Table \ref{tab:bseparams}) over the duration of the BSE simulations (15 Gyr): Kepler-34, -38, -47, -64, and -1647. Kepler-35 experiences a CE for ${\rm Z - 1\sigma_Z}$, and Kepler-453 for ${\rm Z - 2\sigma_Z}$. 

\subsection{Kepler-34}

The binary consists of 1.05\Msun + 1.02\Msun~stars on a highly-eccentric ($e_{\rm bin,0}$ = 0.52), 28-day orbit which changes little in the first $\sim9$ Gyr. The CBP has an initial orbit of 1.09~AU. 

The primary star fills its Roche lobe at $\sim9.6$ Gyr, the binary enters a CE stage and triggers a double-degenerate Supernova event, disrupting the system.

\subsection{Kepler-38}

The binary consists of 0.95\Msun + 0.25\Msun~stars on a slightly-eccentric ($e_{\rm bin,0}$ = 0.1), 19-day orbit which changes little in the first $\sim13$ Gyr. The CBP has an initial orbit of 0.46~AU. The main evolution stages of the binary star and their effect on the CBP are as follows.

\subsubsection{The Binary}

The primary star fills its Roche lobe at $t\sim13.7$ Gyr and the binary enters a primary CE stage. During this stage $\Psi(\aalpha)\approx0.04$ and $\ac\approx26~\Msun / yr$ -- much larger than \aalpha and \aalphao. As a result of the CE, the binary merges as a First Giant Branch star and loses $\sim25-40\%$ of its mass for ${\rm \alpha_{CE} = 0.5}$; here $\beta>\beta_{\rm eject}$ (see Eqn. \ref{eq:ejection}), and the planet should remain bound in a runaway regime (i.e. $\Psi\gg1$) even at pericenter (see Eqn. 37-44, Veras et al. (2011)). The system continues to slowly lose mass by the end of the BSE simulations (15 Gyr), thus \aPCE adiabatically expands by up to a factor of 2. 

For ${\rm \alpha_{CE} = 1.0, 3.0, 5.0, 10.0}$ the binary orbit shrinks by a factor of $4-30$ and its mass decreases by $\sim60\%$; here $\beta<\beta_{\rm eject}$, i.e. planet should be ejected in a runaway regime. For \aCE = 1 and TCP the system experiences a secondary RLOF at $t\sim13.74$ Gyr and coalesces into a First Giant Branch star without mass loss (thus no changes in \aPCE or \ePCE). The binary does not lose mass additional mass by the end of the BSE simulations.

Overall, by 15 Gyr the binary evolves into either a merged Helium or Carbon/Oxygen White Dwarf (HeWD, COWD) or a very close HeWD-MS star binary. 

The orbital reconfiguration of the binary system during the primary CE stage is shown on the left panels of Figure \ref{fig:K38_orbits_reconfig} for ${\rm \alpha_{CE} = 0.5}$ (upper panels) and ${\rm \alpha_{CE} = 1.0, 3.0, 5.0, 10.0}$ (lower panels). The observer is looking along the y-axis, from below the figure. The maximum (red dashed; binary initially at {\bf Eastern} Elongation (EE), CBP initially at phase 0.73) and minimum (red solid; binary at Easter Elongation (EE), CBP at phase 0.08) orbital expansion of the CBP is shown on the right panels of the figure for \aalpha (red), as well as the maximum expansion for \aalphao (green solid). To test the dynamical stability of the CBP at maximum expansion (dashed red), we integrated the system for a thousand planetary orbits. The orbit precesses, but does not exhibit chaotic behavior for the duration of the integration. We leave examination of the long-term behavior and stability of the system for future work.

\begin{figure}
\centering
\epsscale{1.1}
\plotone{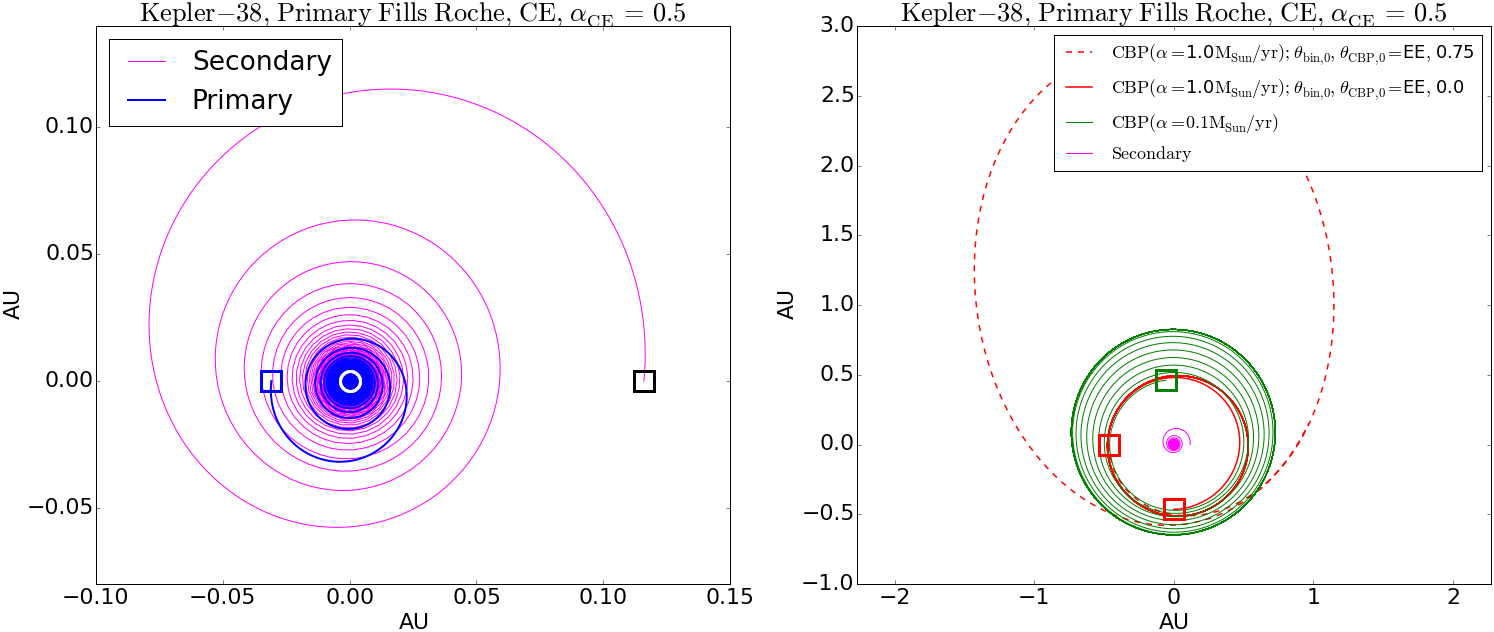}
\plotone{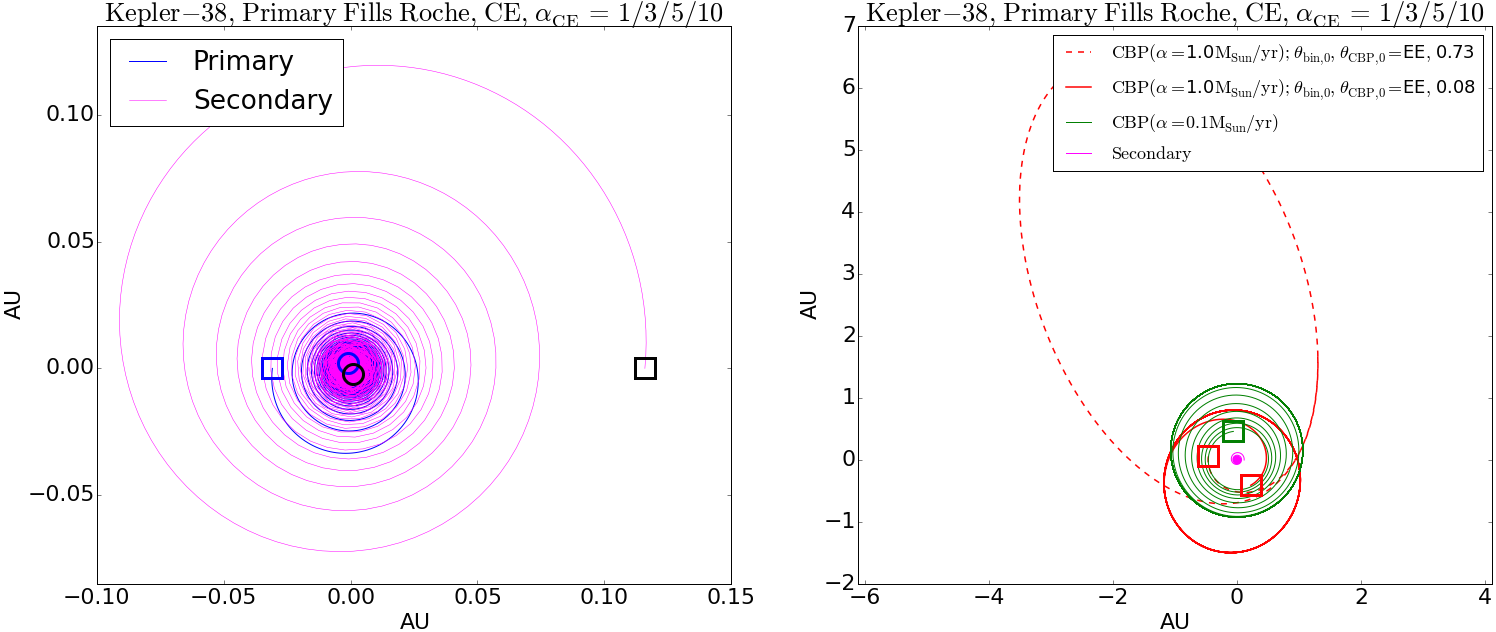}
\caption{Orbital reconfiguration of the Kepler-38 system during the Primary CE stage. The binary orbits are shown on the left panels (blue for the primary star, magenta for the secondary) and the CBP orbits -- on the right (green for \aalphao,  red for \aalpha,~where dashed red corresponds to the largest and the solid red to the smallest achieved orbit). The binary merges for ${\aCE = 0.5}$. The square symbols indicate the initial positions of the two stars and the planet; the circle symbols in the left panels indicate the final positions of the primary (blue) and secondary (black) star. The right panels show the evolution of \ap~from different initial conditions as indicated by the respective squares. 
\label{fig:K38_orbits_reconfig}}
\end{figure}

\subsubsection{The CBP}

On Fig. \ref{fig:K38prim_merger} we show the \aCE = 0.5 CE evolution of \Mprim, \abin (blue, upper panel), and of \ap (middle panels) and \ep (lower panel) for \aalpha (red) and \aalphao (green) respectively; on Fig. \ref{fig:K38prim_nonmerger} we show the corresponding evolution for \aCE = 1/3/5/10. Unlike the case of a planet orbiting a single star, where the evolution of the planet's orbit follows a single path for each \aM (see Fig. \ref{fig:K38_single_star}), the dynamical response of a CBP to a binary undergoing a CE stage depends on the initial configuration of the system, i.e. the phase difference ${\rm \Delta\theta_0\equiv\theta_{CBP,0}-\theta_{bin,0}-\omega_{bin}}$, where ${\rm \theta_{CBP,0},\theta_{bin,0}}$ are the initial true anomaly of the CBP and the binary respectively, and ${\rm \omega_{bin}}$ is the binary's argument of pericenter. This is reminiscent of the importance of the initial true anomaly of a planet orbiting a single star on an eccentric orbit (see \cite{veras11}). The dependence is shown in the second panel of Fig. \ref{fig:K38prim_merger}, where each curve represents the evolution of \ap for ${\rm \Delta\theta_0}$ varying between 0 and 1 (for clarity, the curves here represent every tenth phase where in the simulations the phase difference varies with a step of 0.02).

Since our simulations encompass a discrete set of initial conditions for \aCE, tides (NTCP or TCP for minimum to maximum strength respectively), and ${\rm \Delta\theta_0}$, the true evolution of \ap and \ep would be continuous. We represent this by the hatched regions on the lower panels of Fig. \ref{fig:K38prim_merger} and \ref{fig:K38prim_nonmerger}, where the different hatches correspond to TCP or NTCP. For clarity, we show individual \ap evolutionary tracks only in Fig. \ref{fig:K38prim_merger}; these are replaced with hatched regions in all subsequent figures. We note that the distribution of \aPCE and \ePCE (i.e.~\ap and \ep at the end of the CE phase) is neither uniform nor necessarily normal, as seen from the second panel on Fig. \ref{fig:K38prim_merger}, but instead depends on ${\rm \Delta\theta_0}$. This dependence is demonstrated on Fig. \ref{fig:K38prim_merger_a_vs_phase} for \aCE = 0.5 and on Fig. \ref{fig:K38prim_nonmerger_a_vs_phase} for \aCE = 1/3/5/10, and the corresponding probability distribution functions (PDFs) are shown in Fig. \ref{fig:K38prim_merger_histo} and \ref{fig:K38prim_nonmerger_histo}. The PDFs can have a prominent peak with a long one-sided tail (e.g. upper left panel on the first figure), or be double-peaked with the peaks near the edges (e.g. lower left panel on the first figure); this is a natural consequence of the vaguely sinusoidal dependencies seen in Figures \ref{fig:K38prim_merger_a_vs_phase} and \ref{fig:K38prim_nonmerger_a_vs_phase}. Given the prominent diversity  of the PDFs, the distributions are better represented by their modes than their medians. Thus for completeness we list the mode, median, and min/max range for each \aPCE and \ePCE in Tables \ref{tab:temp_tab_CBP_fast} through 10, but cite the 68\%-range upper and lower bounds on the modes only as the medians may not be appropriate in some cases. Unless otherwise noted, we refer to the modes of \aPCE and \ePCE as the default results. 
 
It is important to note that the osculating orbital elements of a CBP do not describe a closed ellipse during the CE phase and its orbital motion is not Keplerian. Instead, the orbit spirals outwards with time and, as seen from Fig. \ref{fig:K38prim_merger} and \ref{fig:K38prim_nonmerger}, while the binary is losing mass \ap and \ep oscillate together with the oscillations in \ebin (see inset figure, lower panel). Similar behavior was observed in numerical simulations of planets around single stars experiencing mass loss \citep{veras11, adams13} where the planet's orbital elements oscillate on the planet's period due to the different effects of mass-loss at pericenter and apocenter. This is reproduced in our simulations for the binary star itself, i.e. the orbital elements of the secondary star oscillate on the binary period as seen from the inset figure in the lower panel of Fig. \ref{fig:K38prim_merger}. The CBP, initially on a circular orbit quickly gains eccentricity and responds to the binary's oscillations.

Overall, the orbital expansion of the Kepler-38 CBP for \aalphao is fully consistent with the adiabatic approximation, and the planet gains slight eccentricity (see Table \ref{tab:temp_tab_CBP_slow}). The outcome for the case of \aalpha~is diverse -- \aPCE ranges from below the corresponding adiabatic approximation up to 3.8 AU, and \ePCE can reach up to 0.8 (see Table \ref{tab:temp_tab_CBP_fast}). 

\begin{figure}
\centering
\epsscale{1.}
\plotone{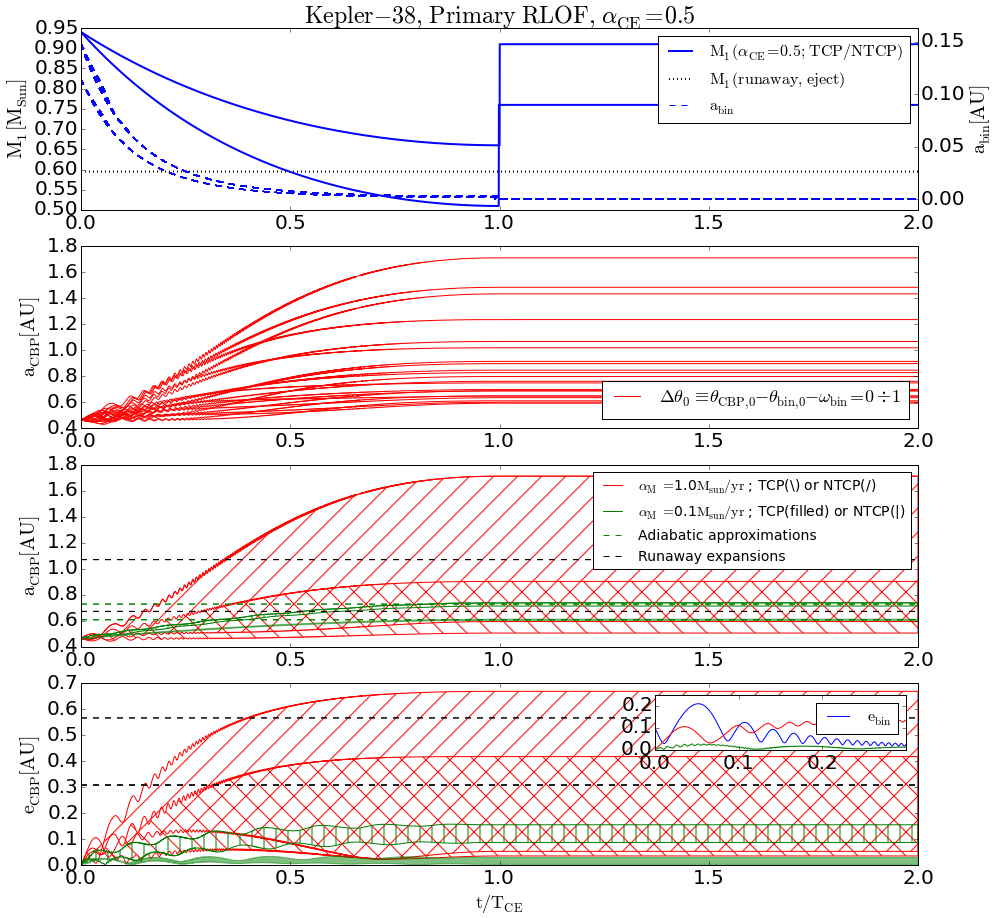}
\caption{Evolution of, from top to bottom, the mass of the Kepler-38 primary star and the binary semi-major axis (blue), \ap, and \ep (red for \aalpha, green for \aalphao) during the primary RLOF and CE for ${\rm \alpha_{CE} = 0.5}$. The binary merges at the end of the CE, i.e. at ${\rm t=T_{CE}}$. The second panel shows the dependence of the evolution of \ap on ${\rm \Delta\theta_0}$, the initial true anomaly difference between the binary star and the CBP at the onset of the CE (${\rm \Delta\theta_0\equiv\theta_{CBP,0}-\theta_{bin,0}-\omega_{bin}}$). Unlike the case of a planet orbiting a single star losing mass, where the planet's orbit follows a single evolutionary track, a CBP orbiting a binary star can experience different evolutionary tracks during a CE phase for different ${\rm \Delta\theta_0}$. The individual evolutionary tracks for \ap and \ep are condensed into hatched regions on the third and fourth panels (and in the subsequent figures) for clarity. The dotted black line in the upper panel represents the critical mass for ejection of the CBP, the dashed lines in the middle and lower panels represent the respective adiabatic and runaway expansions.
\label{fig:K38prim_merger}}
\end{figure}

\begin{figure}
\centering
\plotone{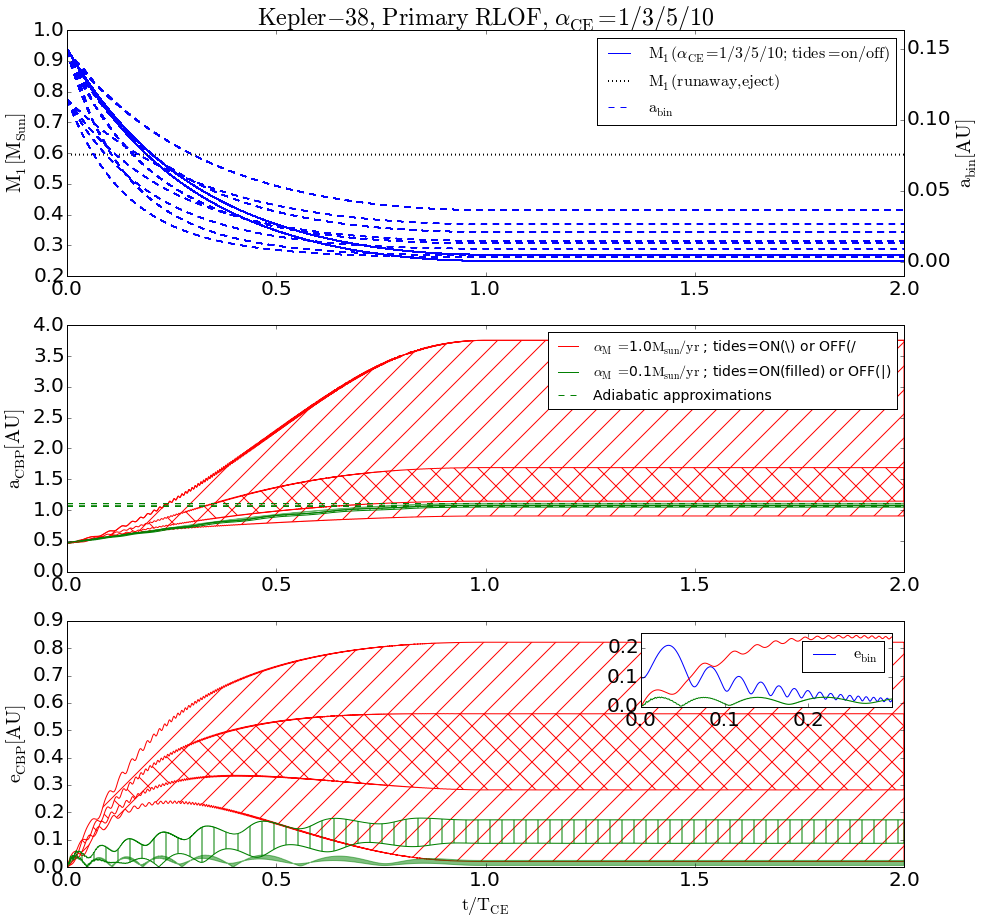}
\caption{Same as Fig. \ref{fig:K38prim_merger} but for \aCE = 1/3/5/10.
\label{fig:K38prim_nonmerger}}
\end{figure}

\begin{figure}
\centering
\plotone{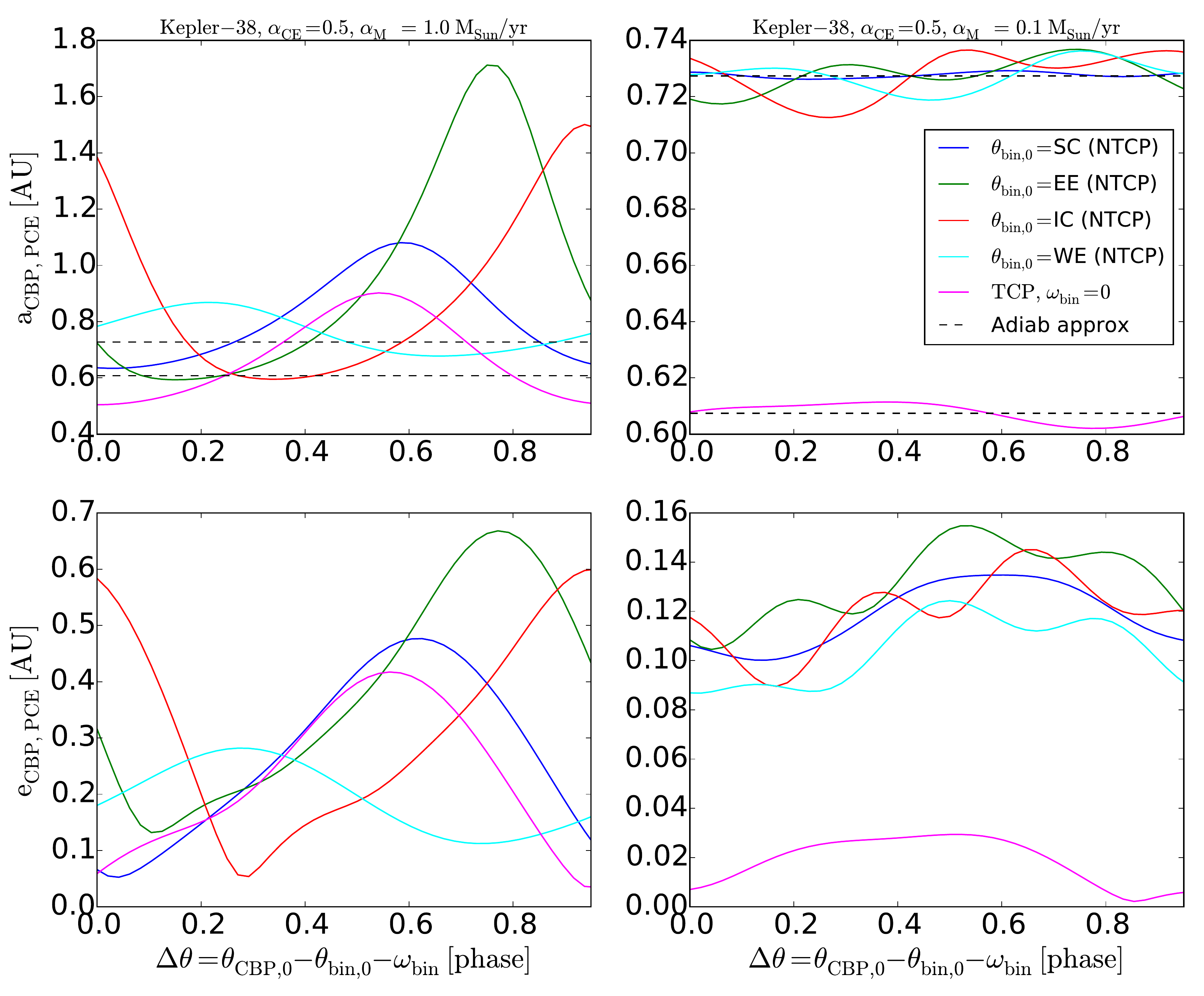}
\caption{Dependence of ${\rm a_{CBP, PCE}}$ (upper) and ${\rm e_{CBP, PCE}}$ (lower) on the initial phase difference between the binary and the CBP $\Delta\theta_0$, and on \aM for Kepler-38, and for \aCE = 0.5. The different colors for NTCP (blue, red, green, cyan) correspond to different initial binary phases, i.e. Eastern and Western Elongation (EE and WE), and Superior and Inferior Conjunction (SC, IC). The case of TCP is represented by a single color (magenta) as the binary is circular at the onset of the RLOF, and thus has a single initial condition (set to inferior conjunction for simplicity). The dashed black lines in the upper panels represent the corresponding adiabatic expansion approximation. As seen from the upper right panel the results from our numerical simulations for \aPCE and for \aalphao are fully consistent with the adiabatic approximation. For \aalpha, the modes of the \aPCE and \ePCE distributions are 0.5/0.6 AU and 0.4/0.15 for TCP/NTCP respectively, and the planet can reach \aPCE of up to 1.7 AU, and \ePCE of up to 0.67.
\label{fig:K38prim_merger_a_vs_phase}}
\end{figure}

\begin{figure}
\centering
\plotone{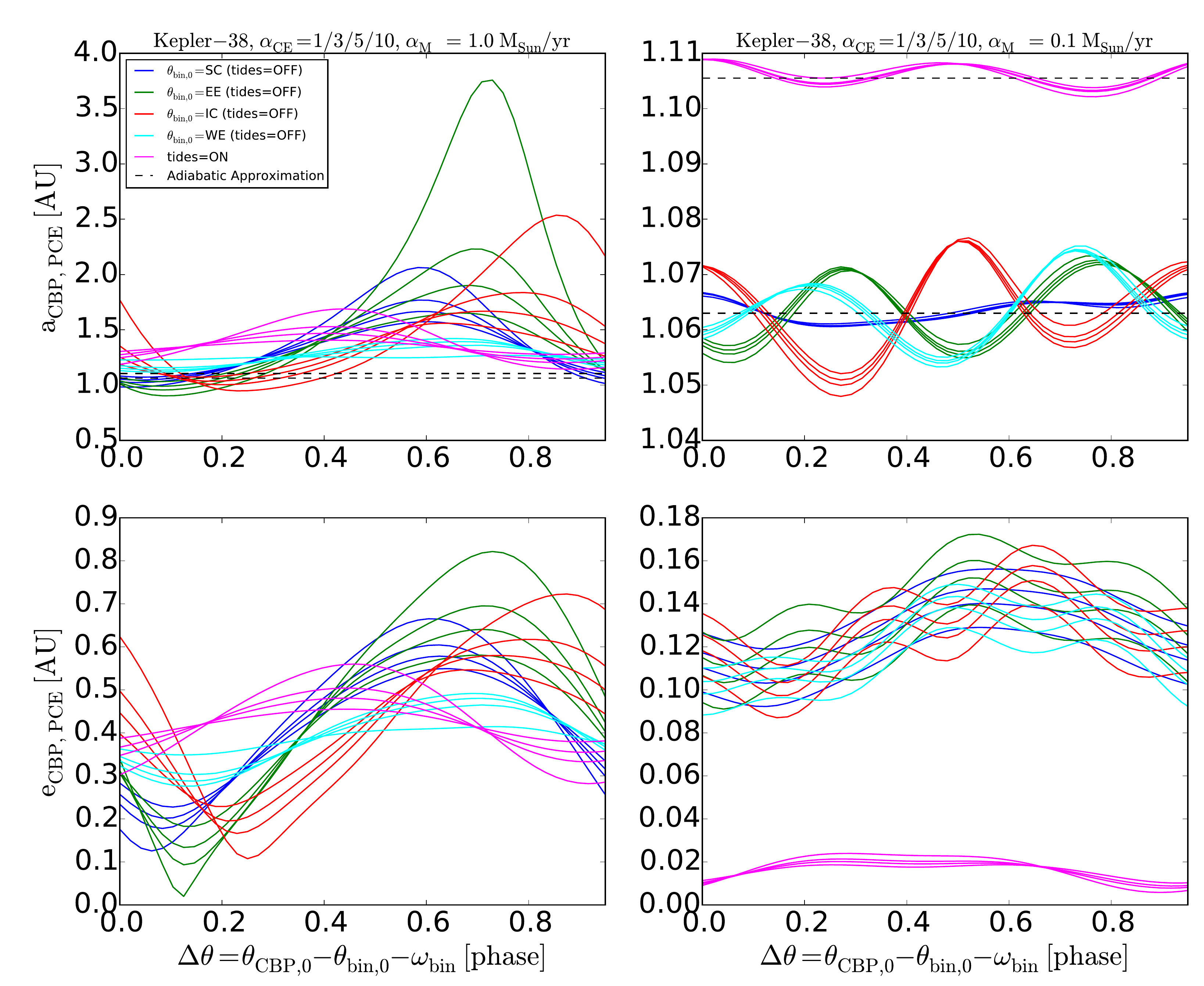}
\caption{Same as Fig. \ref{fig:K38prim_merger_a_vs_phase} but for \aCE = 1/3/5/10. For \aalpha, mode(\aPCE) = 1.4/1.3 AU and mode(\ePCE) = 0.46/0.38 for TCP/NTCP respectively; \aPCE can reach 3.8 AU, and \ePCE can be as high as 0.82.
\label{fig:K38prim_nonmerger_a_vs_phase}}
\end{figure}

\begin{figure}
\centering
\plotone{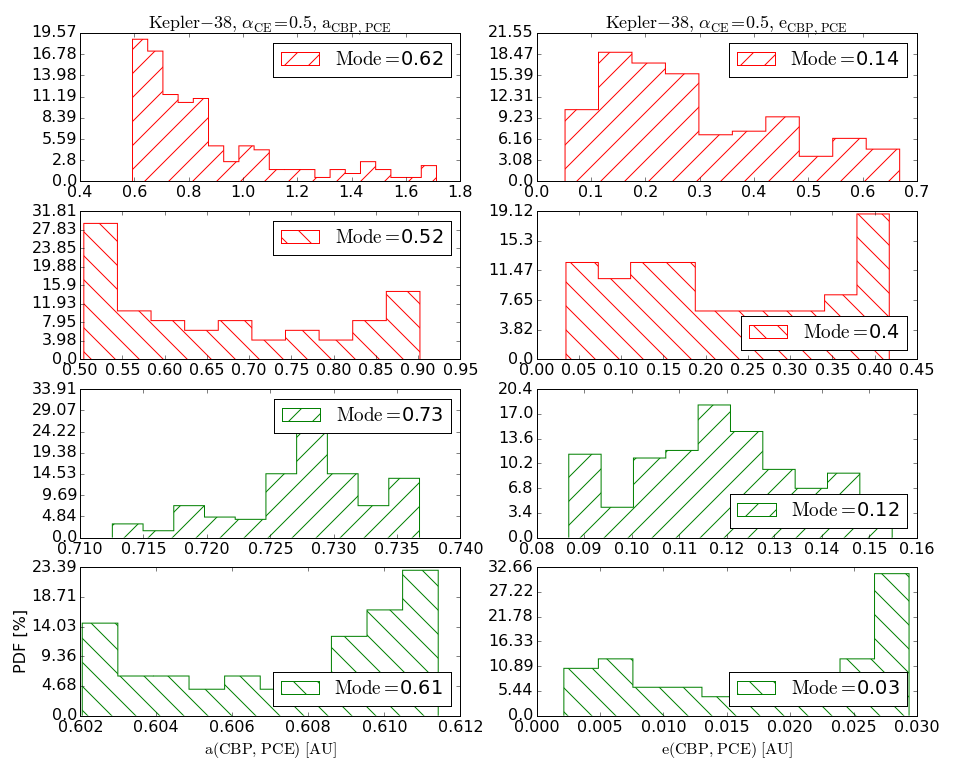}
\caption{Probability Distribution Functions for ${\rm a_{CBP, PCE}}$ (left) and ${\rm e_{CBP, PCE}}$ (right) for Kepler-38 for \aCE = 0.5. Red histograms represent ${\rm \alpha_M = 1~\Msun/yr}$, green histograms represent ${\rm \alpha_M = 0.1~\Msun/yr}$ ($/$ hatch for NTCP, \textbackslash~hatch for TCP).
\label{fig:K38prim_merger_histo}}
\end{figure}

\begin{figure}
\centering
\plotone{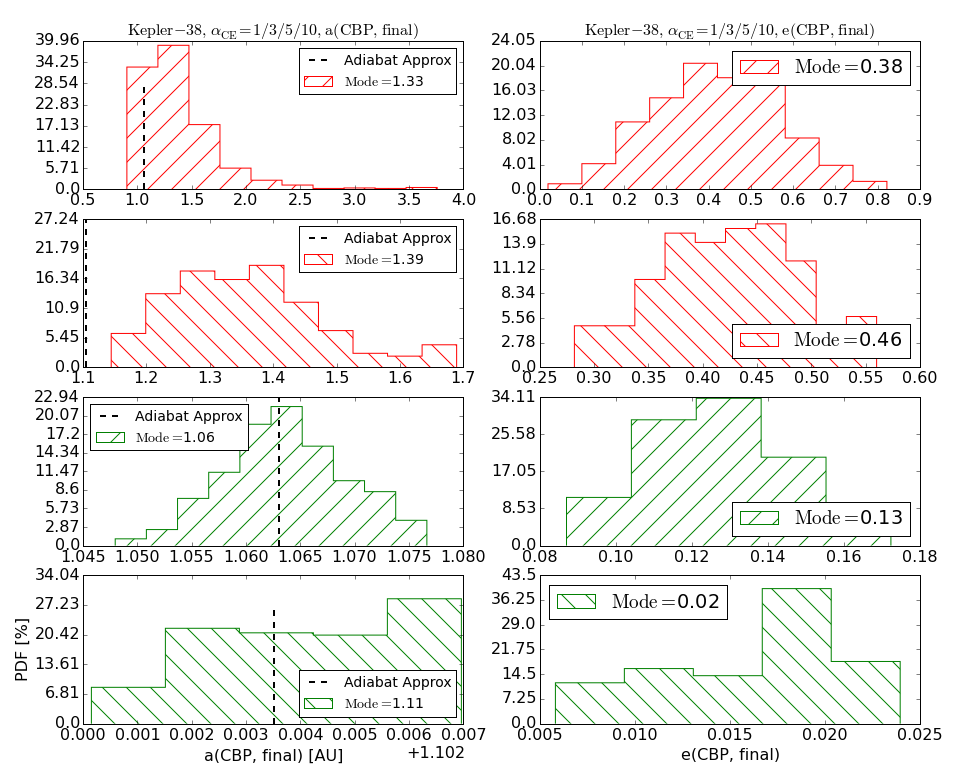}
\caption{Same as Fig. \ref{fig:K38prim_merger_histo} but for \aCE = 1/3/5/10.
\label{fig:K38prim_nonmerger_histo}}
\end{figure}

\subsection{Kepler-47}

The binary consists of 0.96\Msun + 0.34\Msun~stars on a nearly circular ($e_{\rm bin,0}$ = 0.02), 7.45-day orbit which changes little in the first $\sim12$ Gyr. The three CBPs have initial orbits of 0.3~AU, 0.72~AU and 0.99~AU, and masses of 2~${\rm M_\Earth}$, 19~${\rm M_\Earth}$, and 3~${\rm M_\Earth}$ (J.~Orosz, priv.~comm.), from inner to outer respectively. For completeness, we integrate the system for the appropriate CBP masses to take into account planet-planet interactions. The main evolutionary stages are as follows.

\subsubsection{The Binary}

The primary star experiences a RLOF at ${\rm t\sim12}$ Gyr and the binary enters a CE stage. During this stage $\Psi(\aalpha)\approx0.02/0.06/0.1$ and $\ac\approx60/16/10~\Msun / yr$ for planets 1/2/3 respectively. For \aalpha~ the time of CE mass loss (\tCE) is comparable to the orbital periods of planets 2 and 3, i.e. ${\rm T_{CE}/P_{CBP}\sim0.5-2}$, indicating that these two CBPs evolve in the transition regime. 

As a result of the primary CE, for \aCE = 0.5/1 the binary merges as a First Giant Branch star and loses $\sim15-40\%$ of its mass; $\beta<\beta_{\rm eject}$ and the CBPs should remain bound even in a runaway regime. From the end of the CE to the end of the BSE simulations, the system slowly loses  50\% of its mass, and \aPCE of those planets that remain long-term stable after the CE (discussed below) expand adiabatically by a factor of 2.

For \aCE = 3/5/10, the binary evolves into a HeWD-MS star pair, loses $56\%$ of its mass and \abin decreases by a factor of $4-10$; here $\beta>\beta_{\rm eject}$ and the CBPs should be ejected in a runaway regime. The system does not experience further mass loss by the end of the BSE simulations.  

\subsubsection{The CBPs}

On Fig. \ref{fig:K47prim_merger} and \ref{fig:K47prim_nomerger} we show the evolution of \ap and \ep for the three CBPs (and for \aalpha and \aalphao) caused by the primary RLOF and CE. For simplicity here and for the rest of the \kepler systems we do not show the evolution of \Mprim~and \abin since it is qualitatively very similar to the case of Kepler-38. We note that, unlike the corresponding figures for Kepler-38, here the colors represent the three planets (magenta, green and red for Planets 1, 2 and 3 respectively), and the different panels represent the evolution of \ap and \ep for \aalpha (upper and middle panels), and for \aalphao (lower panels). 

\begin{figure}
\centering
\plotone{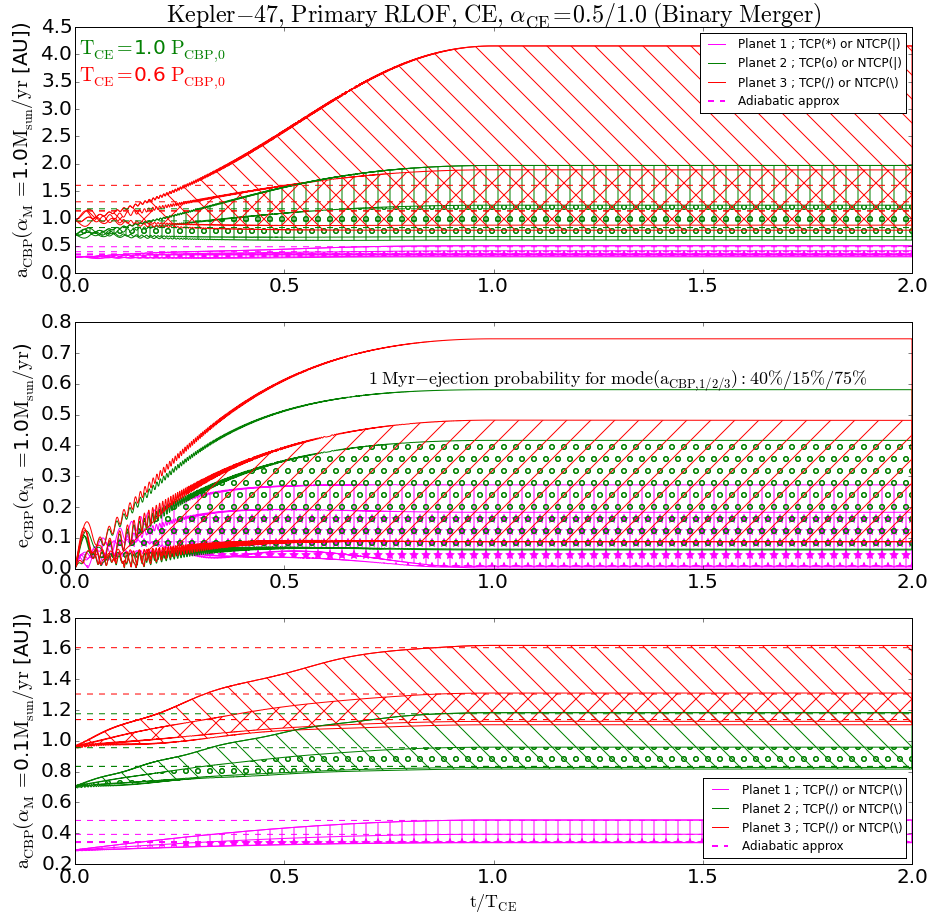}
\caption{Evolution of Kepler-47 CBP's \ap and \ep during the primary RLOF and CE for ${\rm \alpha_{CE} = 0.5/1.0}$ (binary merger), for \aalpha (upper two panels) and for \aalphao (lower panel), where the colors represent planets 1 (magenta), 2 (green) and 3 (red). The dashed lines represent the corresponding adiabatic and runaway expansions. The orbital periods of Planets 2 and 3 are comparable to the mass-loss timescale of the CE (${\rm T_{CE}}$) and evolve in the transition regime for \aalpha. All 3 planets evolve adiabatically for \aalphao. PCE planet-planet interactions result in 40\%/15\%/75\% ejection probability for the modes of ${\rm a_{CBP,1/2/3}}$ after 1 Myr. For \aalphao the three \ap evolve adiabatically.
\label{fig:K47prim_merger}}
\end{figure}

\begin{figure}
\centering
\plotone{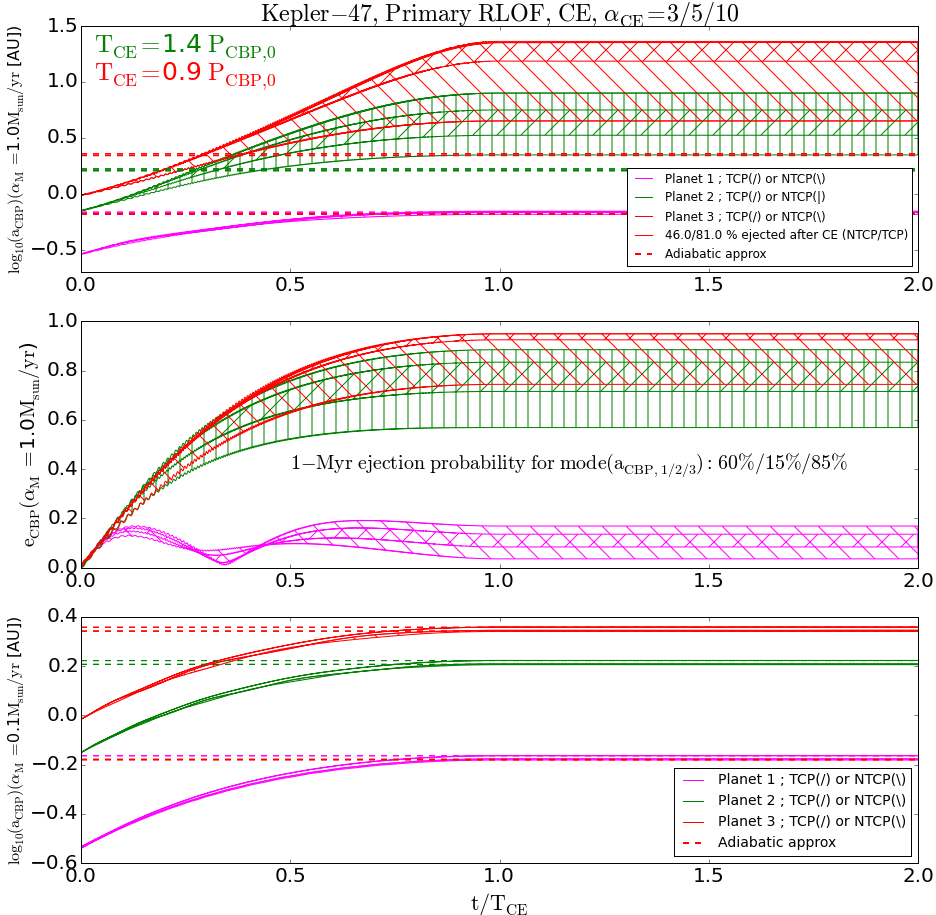}
\caption{Same as Fig. \ref{fig:K47prim_merger} but for \aCE = 3/5/10. The orbital periods of Planets 2 and 3 are comparable to the CE mass-loss timescale ${\rm T_{CE}}$ and evolve in the transition regime for \aalpha; planet 3 is ejected during the CE in 46\%/81\% of the simulations for NTCP/TCP respectively. For \aalpha, planets 1/2/3 are ejected in 60\%/15\%/85\% after 1-Myr. For \aalphao their orbits evolve adiabatically and do not experience ejections. 
\label{fig:K47prim_nomerger}}
\end{figure}

As seen from the lower panels on Fig. \ref{fig:K47prim_merger} and \ref{fig:K47prim_nomerger}, the orbital expansion of all three of Kepler-47's CBPs is fully consistent with the adiabatic approximation for \aalphao, and the planets gain slight eccentricity (not shown in the figures, but listed in Table \ref{tab:temp_tab_CBP_slow}). The orbital evolution of Planets 2 and 3 for the primary RLOF is much richer for \aalpha, as shown on the upper and middle panels of Fig. \ref{fig:K47prim_merger} and \ref{fig:K47prim_nomerger} (see also Table \ref{tab:temp_tab_CBP_fast}). The planets gain high eccentricities and for \aCE = 3/5/10 may even become ejected during the CE stage, where given the complexity, uncertainties and large parameters space of the evolution of the binary, and of the CBP here and throughout the paper we define ejection as $\ep > 0.95$. Specifically, planet 3 becomes unbound in 46\% of the NTCP simulations, and in 81\% of the TCP simulations. 

We note that during the primary CE stage, for the case of \aalpha~the orbit of the inner CBP grows adiabatically while those of the outer do not (see upper panel of Fig. \ref{fig:K47prim_nomerger}). This is because the orbital period of the inner planet is much shorter than the duration of \tCE, while those of the outer two are comparable. Thus a multiplanet CB system can experience two regimes of planetary orbital evolution during the same CE stage. 

The dependence of \aPCE and \ePCE of Planet 2 on ${\rm \Delta\theta_0}$ is shown on Fig. \ref{fig:K47prim_merger_a_vs_phase_2} and \ref{fig:K47prim_nonmerger_a_vs_phase_2}. The planet's orbit evolves adiabatically for \aalphao (upper right panels), non-adiabatically for \aalpha, and remains bound during the CE in both cases, regardless of ${\rm \Delta\theta_0}$.

\begin{figure}
\centering
\plotone{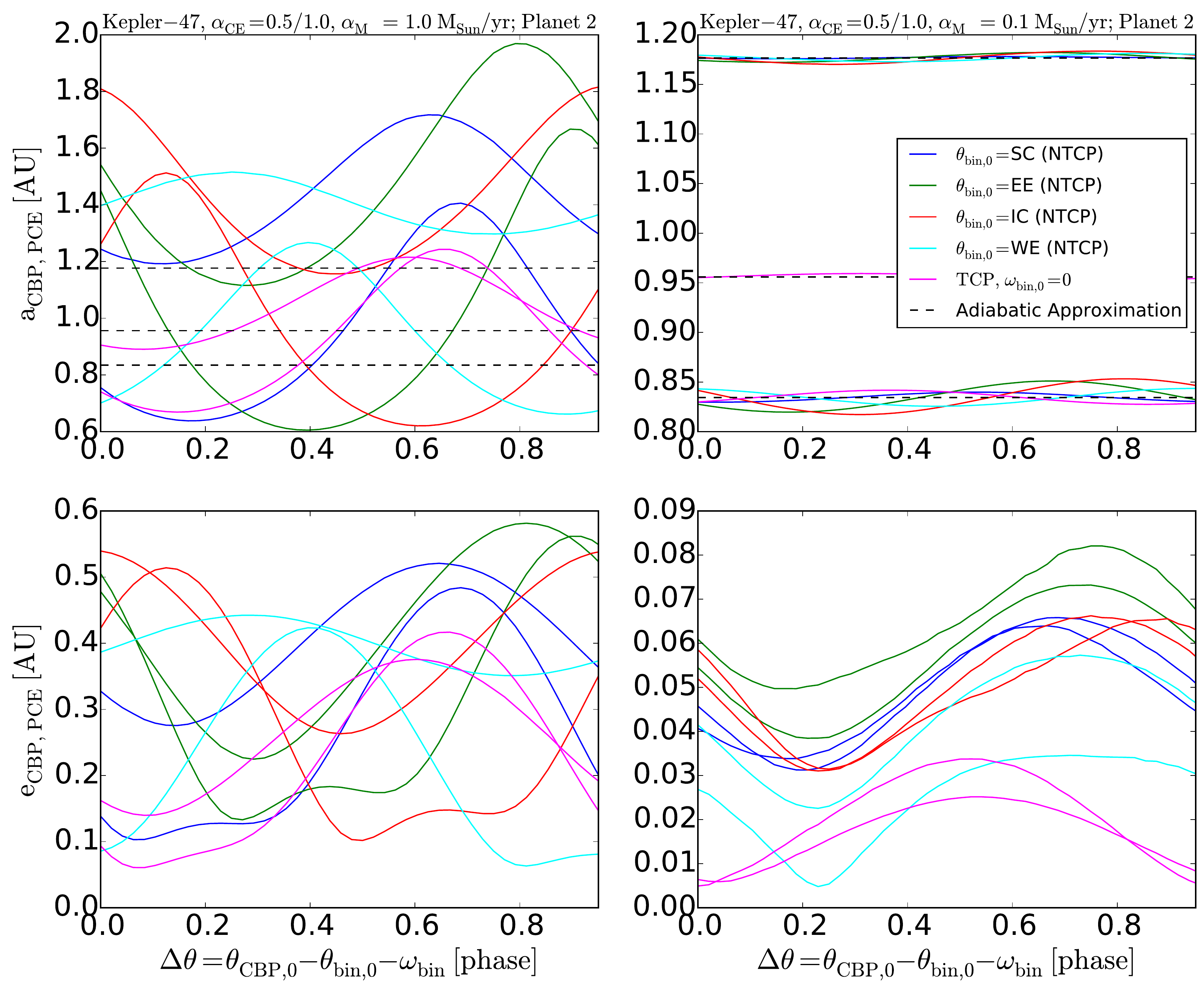}
\caption{Same as Fig. \ref{fig:K38prim_merger_a_vs_phase} but for Kepler-47, Planet 2 and for \aCE = 0.5/1. For \aalpha, mode(\aPCE) = 0.6-1.33 AU and mode(\ePCE) = 0.08-0.42, depending on \aCE and tidal evolution, and can reach 1.97 AU and 0.58 eccentricity. The results from our numerical simulations for \aPCE and for \aalphao are fully consistent with the adiabatic approximation. The planet remains bound regardless of $\Delta\theta_0$. 
\label{fig:K47prim_merger_a_vs_phase_2}}
\end{figure}

\begin{figure}
\centering
\plotone{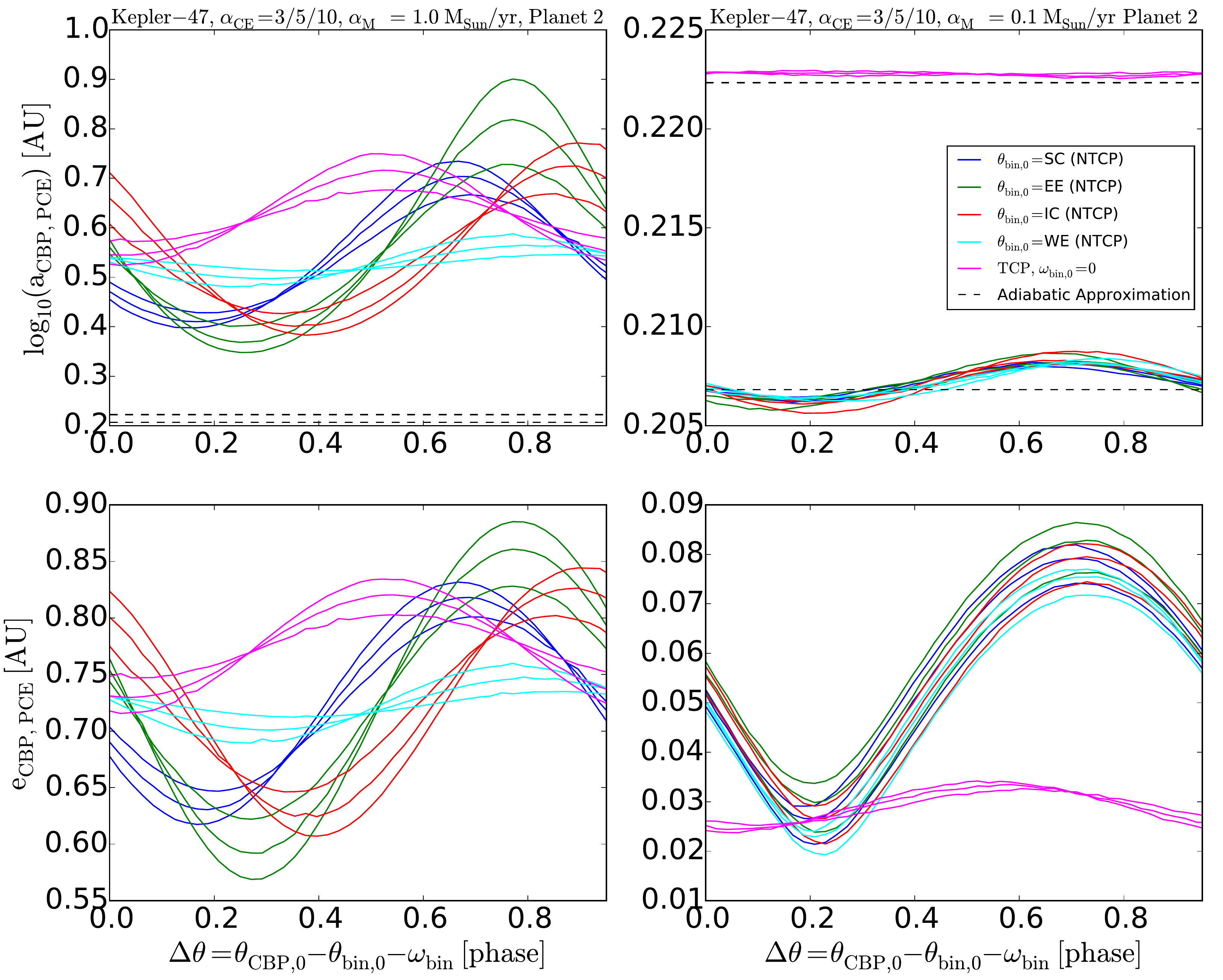}
\caption{Same as Fig. \ref{fig:K47prim_merger_a_vs_phase_2} but for \aCE = 3/5/10. For \aalpha, mode(\aPCE) = 3.68/3.23 AU, mode(\ePCE) = 0.71/0.75 for TCP/NTCP, \aPCE can reach 7.95 AU and \ePCE can be as high as 0.89. The orbit of the planet evolves adiabatically for \aalphao, and non-adiabatically for \aalpha. The planet remains bound in both cases, regardless of $\Delta\theta_0$. 
\label{fig:K47prim_nonmerger_a_vs_phase_2}}
\end{figure}

Even if an ejection does not occur during the primary CE phase, \aPCE and \ePCE of all three CBPs for \aalpha~are such that the planets will continue to interact with each other after the end of the CE in such a way that further ejections/collisions are possible. To study the long-term evolution of the system, we integrated the equations of motion for all five bodies (two stars and three planets) for 1 Myr, using the planets' mode(\aPCE) and mode(\ePCE), and randomizing their initial arguments of pericenter and true anomalies. Any planet that achieves $\ep > 0.95$ is marked as ejected and removed from the simulations, and collision is defined as planet-planet separation smaller than ${\rm d_{min} = R_{P,i} + R_{P,j}}$.

Overall, Planet 2 dominates the 1-Myr dynamical evolution and has the highest probability of remaining bound to the system. For \aCE = 0.5/1 the modes of ${\rm a_{CBP, 1/2/3}}$ have 40\%/15\%/75\% ejection probability in 1 Myr. Where Planet 2 remains bound, it's orbital elements remain mostly within the respective 68\%-range of mode(${\rm a_{CBP, PCE, 2}}$) and mode(${\rm e_{CBP, PCE, 2}}$)\footnote{Except in one simulations where planet 1 is ejected, planet 2 migrates to a 5.67 AU, 0.71-eccentricity orbit, and planet 3 migrates to a 0.55 AU, 0.66-eccentricity orbit.}. 

Here the modes of ${\rm a_{CBP, 1/2/3}}$ have 60\%/15\%/85\% ejection probability in 1 Myr, and the orbital elements of Planet 2 again remain mostly within the 68\%-range of mode(${\rm a_{CBP, PCE, 2}}$) and mode(${\rm e_{CBP, PCE, 2}}$).

\subsection{Kepler-64}

The binary consists of 1.53\Msun + 0.41\Msun~stars on a slightly-eccentric ($e_{\rm bin,0}$ = 0.22), 20-day orbit which changes little in the first $\sim3$ Gyr; the CBP has an initial orbit of 0.65~AU. The systems evolves as follows.  

\subsubsection{The binary}

The primary star fills its Roche lobe at ${\rm t\sim3.02}$ Gyr and the binary enters a primary CE stage. During this stage $\Psi(\aalpha)\approx0.03$ and ${\rm \ac\approx32~\Msun / yr}$ -- much larger than \aalpha and \aalphao. However, for \aalpha~the time of mass loss (${\rm T_{CE}}$) is comparable to the orbital period of the CBP, i.e. ${\rm T_{CE}/P_{CBP}\sim1.1-3.4}$, depending on \aCE. Thus while the CBP is theoretically in the adiabatic regime, the comparable timescales result in a non-adiabatic dynamical response. 

The binary merges as a First Giant Branch star and loses $\sim20-58\%$ of its mass for \aCE = 0.5/1. In this regime $\beta>\beta_{\rm eject}$ for \aCE = 0.5, and for \aCE = 1, TCP and the planet should remain bound in a runaway regime even at pericenter; $\beta<\beta_{\rm eject}$ for \aCE = 1.0 and NTCP, indicating a potential ejection of the CBP in a runaway regime. For \aCE = 0.5 and 1 the system continues to slowly lose mass by the end of the BSE simulations (15 Gyr), thus \aPCE expands adiabatically by up to a factor of 3. 

For ${\rm \alpha_{CE} = 3.0, 5.0, 10.0}$ the binary shrinks by a factor of $5-25$ and its mass decreases by $\sim65\%$; here $\beta<\beta_{\rm eject}$ and the CBP should be ejected in a runaway regime. For \aCE = 3 and TCP the system experiences a secondary RLOF at ${\rm t\sim6.2}$ Gyr and coalesces without mass loss (no changes in \ap or \ep). Except for \aCE = 3 and TCP, the binary does not lose mass after the primary CE.

The binary reaches 15 Gyr as either a merged COWD or a very close HeWD-MS star binary. 

\subsubsection{The CBP}

On Fig. \ref{fig:K64prim} we show the evolution of \ap and \ep for \aalpha (red) and \aalphao (green) respectively for \aCE = 0.5/1 (upper two panels) which result in a binary merger, and for \aCE = 3/5/10 (lower two panels). The dependence of \aPCE and \ePCE on the initial phase difference between the binary and the CBP (${\rm \Delta\theta_0}$) is shown on Fig. \ref{fig:K64prim_merger_a_vs_phase} for \aCE = 0.5/1.0 and on Fig. \ref{fig:K64prim_nonmerger_a_vs_phase} for \aCE = 3/5/10. 

Similar to Kepler-38, the orbital expansion of Kepler-64's CBP is fully consistent with the adiabatic approximation for \aalphao, and again the planet gains slight eccentricity (Table \ref{tab:temp_tab_CBP_slow_cont}). The outcome for the case of \aalpha~is quite diverse. Depending on ${\rm \Delta\theta_0}$, the planet can gain very high eccentricity, and even become ejected ($\ep > 0.95$). This is shown as missing points in Fig. \ref{fig:K64prim_merger_a_vs_phase} and \ref{fig:K64prim_nonmerger_a_vs_phase}. The Kepler-64 CBP can reach such eccentricities in $\sim4\%$ of our simulations for \aalpha~and \aCE = 3/5/10. Interestingly, for \aCE = 0.5 and TCP, \ap can decrease during the CE and Mode(\aPCE) = ${\rm 0.6^{+0.6}_{-0.0}~[AU]}$ is smaller than the initial orbit ${\rm a_{CBP,0} = 0.65~AU}$, though the mode's $68\%$-range is significant. On the other side of the spectrum, \ap can reach up to $\sim20$ AU for \aCE=3/5/10 and NTCP.

\begin{figure}
\centering
\plotone{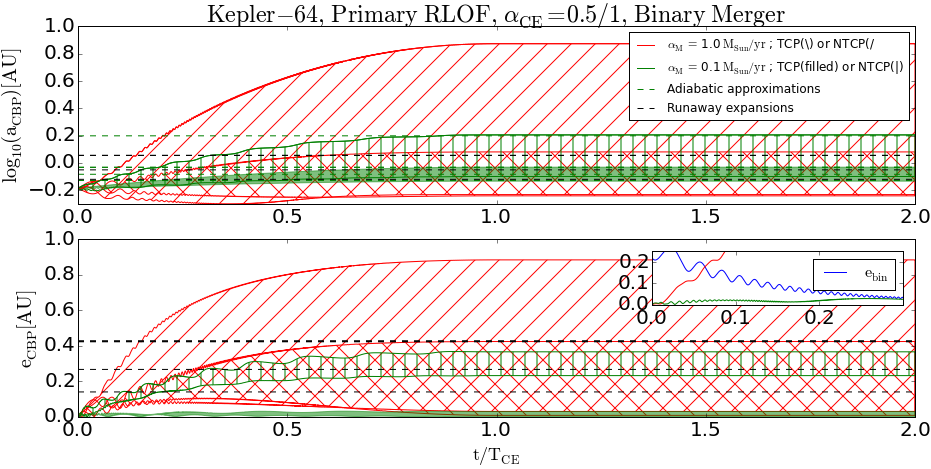}
\plotone{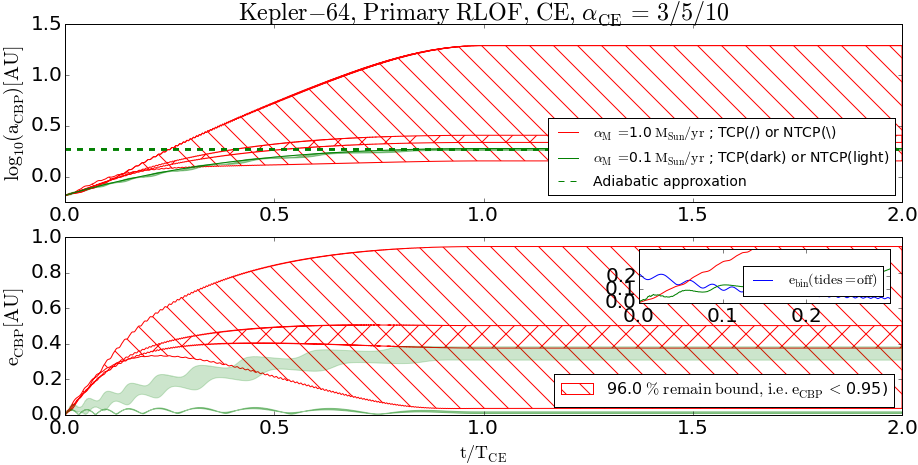}
\caption{Evolution of Kepler-64 CBP's \ap and \ep during the primary RLOF and CE for ${\rm \alpha_{CE} = 0.5,1.0}$ (upper two panels) and for ${\rm \alpha_{CE} = 3, 5, 10}$ (lower two panels), where red color indicates \aalpha, and green color indicates \aalphao. The binary merges at the end of the CE for ${\rm \alpha_{CE} = 0.5,1.0}$. The dashed lines represent the corresponding adiabatic and runaway expansions. For the case of \aalpha and NTCP, $\sim4\%$ of the simulations result in ejection of the CBP (defined as $\ep > 0.95$).
\label{fig:K64prim}}
\end{figure}

\begin{figure}
\centering
\plotone{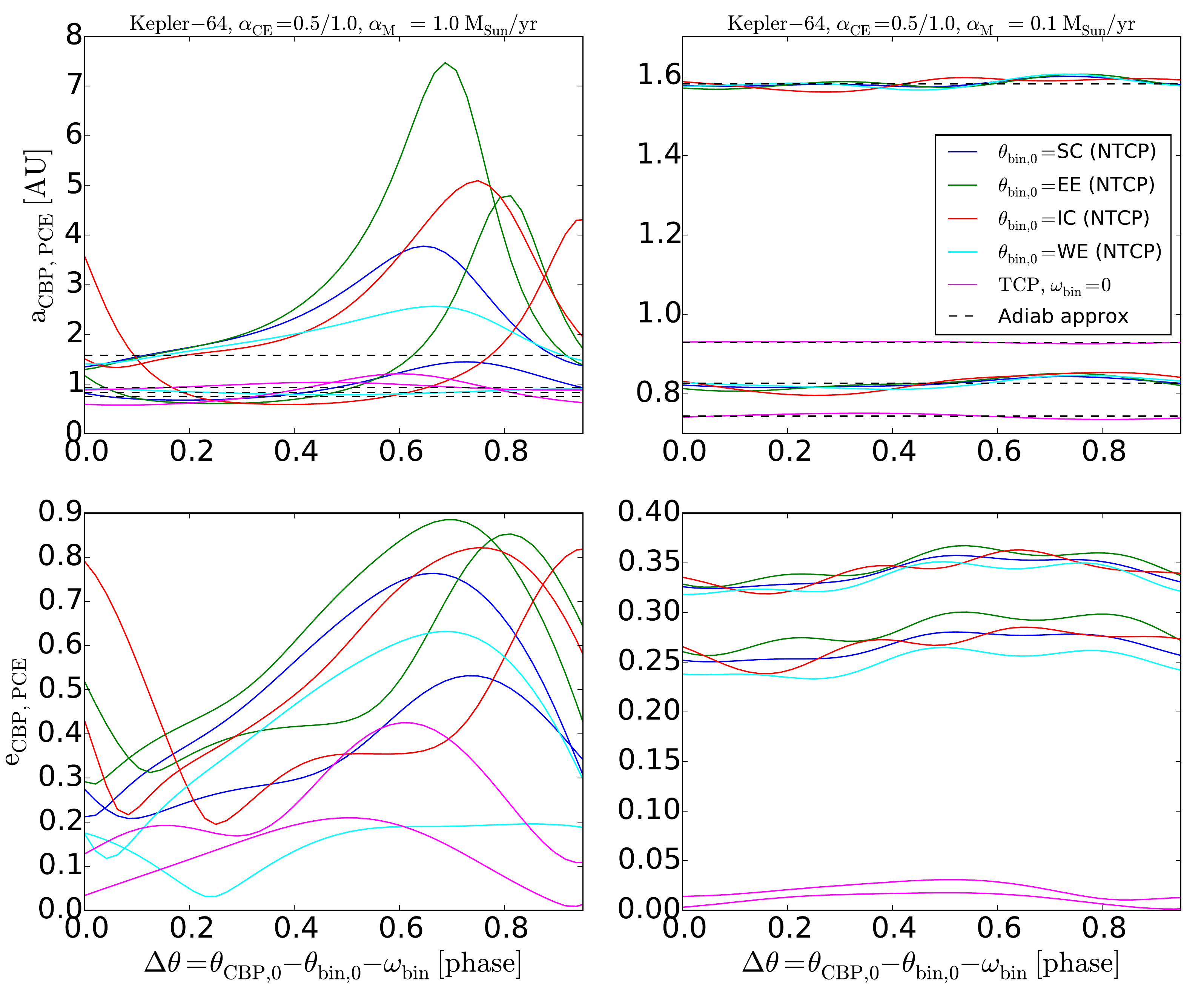}
\caption{Same as Fig. \ref{fig:K38prim_merger_a_vs_phase} but for Kepler-64 and \aCE = 0.5/1. For \aalpha, mode(\aPCE) = 0.6-1.4 AU and mode(\ePCE) = 0.16-0.62 depending on \aCE and tidal evolution; \aPCE can reach 7.5 AU, and \ePCE can reach 0.85. The results from our numerical simulations for \aPCE and for \aalphao are fully consistent with the adiabatic approximation. 
\label{fig:K64prim_merger_a_vs_phase}}
\end{figure}

\begin{figure}
\centering
\plotone{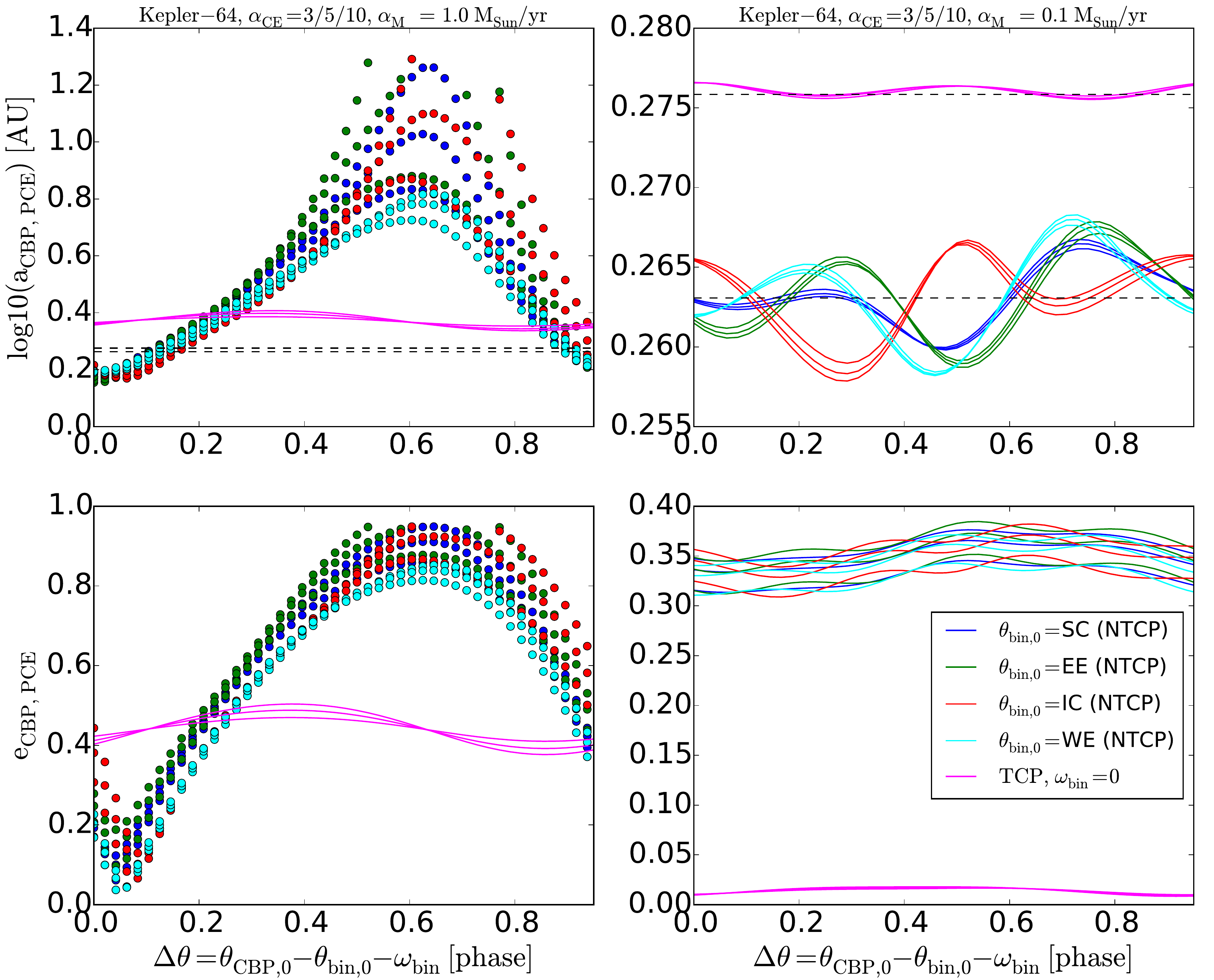}
\caption{Same as Fig. \ref{fig:K64prim_merger_a_vs_phase} but for \aCE = 3/5/10. Here mode(\aPCE) = 2.4/1.6 AU and mode(\ePCE) = 0.49/0.8 for NTCP/TCP,  max(\aPCE) = 19.6 AU and max(\ePCE) = 0.95. The planet is ejected ($\ep > 0.95$) in 4\% of the simulations, indicated by the missing $\Delta\theta_0$ coverage for the green and red colors in the left panels.
\label{fig:K64prim_nonmerger_a_vs_phase}}
\end{figure}

\subsection{Kepler-1647}

The initial binary consists of 1.22\Msun + 0.97\Msun~stars on a slightly eccentric ($e_{\rm bin,0}$ = 0.15), 11-day orbit which changes little in the first $\sim5$ Gyr. The CBP has an initial orbit of 2.7~AU. The main stages of the evolution of the system are as follows. 

\subsubsection{The Binary}

The primary star fills its Roche lobe at ${\rm t\sim5.4}$ Gyr and the binary enters a primary CE stage. During this stage $\Psi(\aalpha)\approx0.2$ and ${\rm \ac\approx4.6~\Msun / yr}$ -- comparable to \aalpha. Additionally, for NTCP and \aalphao~the time of mass loss (${\rm T_{CE}}$) is comparable to the orbital period of the CBP, i.e. ${\rm T_{CE}/P_{CBP}\sim1.3/3.0}$ for \aCE = 0.5/1.0 respectively. Thus the orbit of the CBP evolves in the transition regime. 

Compared to the other \kepler CBP systems, the binary star of Kepler-1647 experiences the richest evolution. For \aCE = 0.5/1 the binary merges as a First Giant Branch star and loses $\sim15-40\%$ of its mass. In this regime $\beta>\beta_{\rm eject}$ and the planet should remain bound in a runaway regime even at pericenter. The system continues to slowly lose mass by the end of the BSE simulations (15 Gyr), thus \aPCE further expands (adiabatically) by up to a factor of 3. 

For ${\rm \alpha_{CE} = 3/5/10}$ the binary shrinks by a factor of $2-7$ into a HeWD-MS star binary and its mass decreases by $\sim45\%$, close to the critical mass loss for runaway ejection (0.5). Here $\beta<\beta_{\rm eject}$ and the CBP should remain bound even in a runaway regime. 

The binary experiences a secondary RLOF and CE for \aCE = 3/5/10 (see Table \ref{tab:temp_tab_EB} for details), and its subsequent evolution can follow several paths. By the end of the BSE simulations, the binary: a) merges without mass loss (i.e.~no changes in \aPCE and \ePCE) into a First Giant Branch for \aCE = 3/5 and TCP, which by 15 Gyr evolves into a COWD; b) evolves into a PCE WD-WD binary (with mass loss, thus \aPCE and \ePCE changes) for \aCE = 5 (and for NTCP) and for \aCE = 10 (for both TCP and NTCP) which can experience a third and final RLOF triggering a SN explosion (for \aCE = 5/10 and TCP).

\subsubsection{The CBP}

The corresponding evolution of \ap and \ep for the primary RLOF, and for \aalpha and \aalphao is shown on Fig. \ref{fig:K1647prim}. Unlike the previous systems, the \aalphao case for Kepler-1647 represents an adiabatic evolution only for the TCP. For \aalphao and NTCP, where \tCE and ${\rm P_{CBP}}$ are comparable for all \aCE, the CBP's orbit evolves in the transition regime (as it does for \aalpha). 

Specifically, for \aalphao and NTCP the CBP reaches sufficiently high eccentricities ($\ep > 0.95$) in $\sim5-25\%$ of the simulations to be ejected from the system; for \aalpha the planet is ejected in $\sim45-55\%$ of the simulations. In addition, for \aCE = 0.5/1 and NTCP, and for \aCE = 3/5/10 and NTCP, \ap can decrease during the primary CE phase and the corresponding mode of \aPCE (1.6-2.3 AU, depending on \aCE, with a wide ${68\%}$-range) is smaller than the initial semi-major axis of the CBP (2.71 AU). Some of the simulations produce the opposite result, namely orbital expansion by up to a factor of $\sim20$ such that \aPCE reaches $\sim50$ AU. The corresponding dependence of \aPCE and \ePCE on $\Delta\theta_0$ are shown in Fig. \ref{fig:K1647prim_merger_a_vs_phase} and \ref{fig:K1647prim_nonmerger_a_vs_phase}, where the missing $\Delta\theta_0$ phase coverage in all panels represent planet ejection. 

\begin{figure}
\centering
\epsscale{1.1}
\plotone{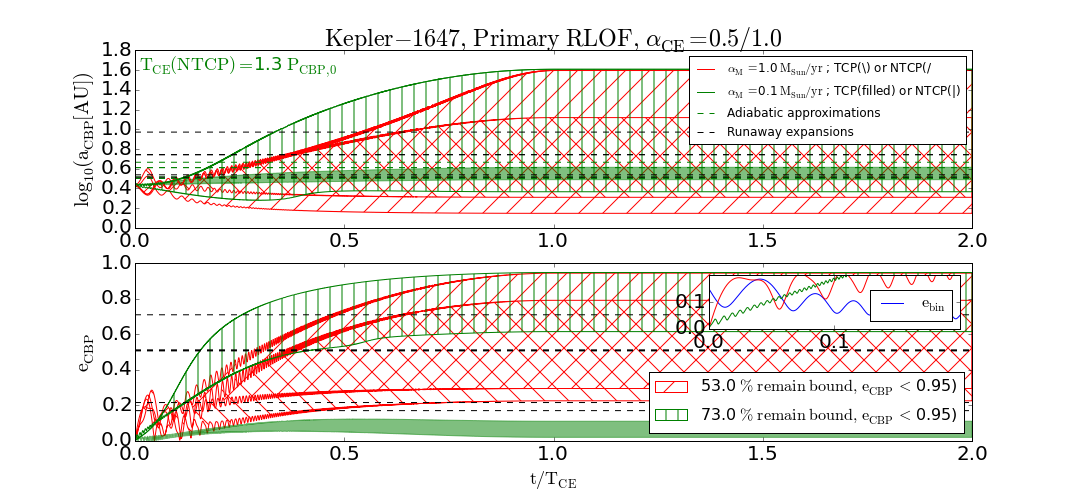}
\plotone{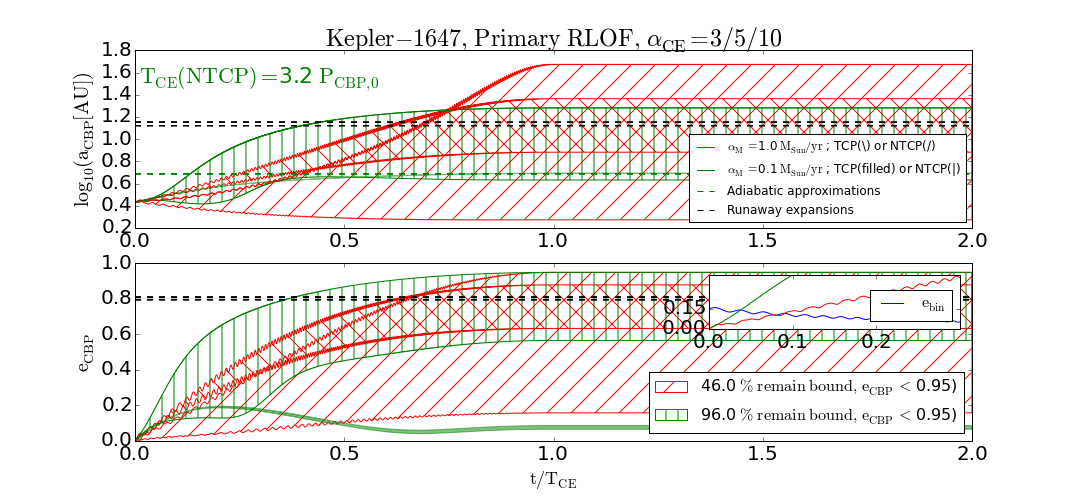}
\caption{Evolution of Kepler-1647 CBP's \ap and \ep during the primary RLOF and CE for ${\rm \alpha_{CE} = 0.5/1}$ (upper two panels) and for ${\rm \alpha_{CE} = 3/5/10}$ (lower two panels), where red color indicates \aalpha, and green color indicates \aalphao. The binary merges at the end of the CE (i.e. ${\rm t = T_{CE}}$) for ${\rm \alpha_{CE} = 0.5/1}$. The dashed lines represent the corresponding adiabatic and runaway expansions. The CBP is ejected ($\ep > 0.95$) in $\sim5-50\%$ of the simulations depending on \aCE and the tidal evolution, for both \aalpha and \aalphao. The latter case is non-adiabatic as ${\rm T_{CE}}$ is comparable to the period of the CBP at the start of the simulations (${\rm P_{CBP,0}}$) and the CBP can expands to large orbits, gaining very high eccentricities. 
\label{fig:K1647prim}}
\end{figure}

\begin{figure}
\centering
\plotone{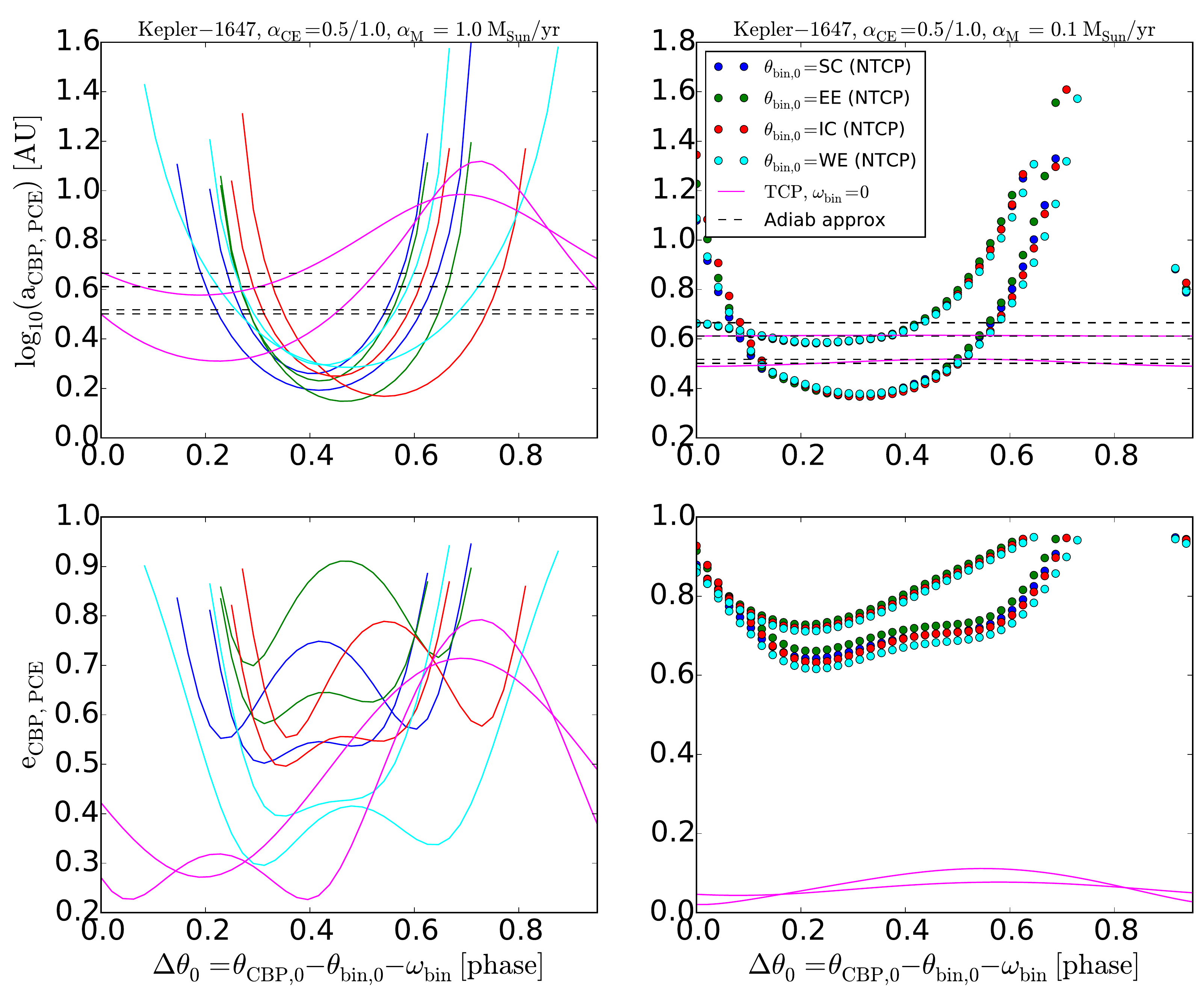}
\caption{Same as Fig. \ref{fig:K38prim_merger_a_vs_phase} but for the primary CE stage of Kepler-1647 and \aCE = 0.5/1. Here mode(\aPCE) = 1.6-3.9 AU, mode(\ePCE) = 0.26-0.72 depending on \aCE and tidal evolution, max(\aPCE) = 37.5 AU and max(\ePCE) = 0.95. The planet is ejected in 47\%/23\% of the \aalpha/\aalphao~simulations, indicated by the missing $\Delta\theta_0$ phase coverage for all but the magenta curves in the left panels and  points in the right panels.
\label{fig:K1647prim_merger_a_vs_phase}}
\end{figure}

\begin{figure}
\centering
\plotone{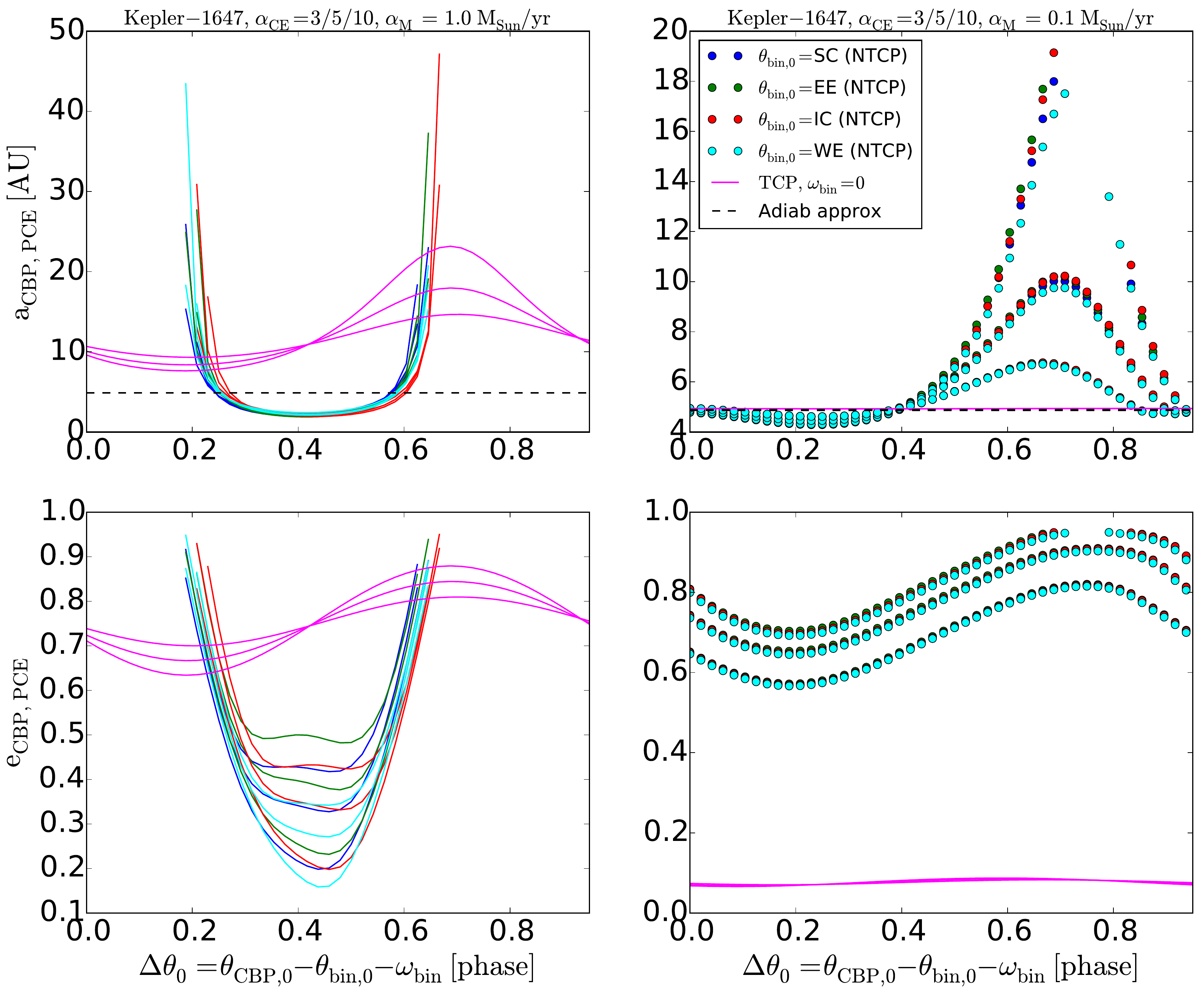}
\caption{Same as Fig. \ref{fig:K1647prim_merger_a_vs_phase} but for \aCE = 3/5/10. Mode(\aPCE) = 9.6/2.3 AU, mode(\ePCE) = 0.7/0.4 for NTCP/TCP, max(\aPCE) = 47.1 AU and max(\ePCE) = 0.95. The planet is ejected in 54\%/4\% of the \aalpha/\aalphao~simulations. 
\label{fig:K1647prim_nonmerger_a_vs_phase}}
\end{figure}

Using the respective ${\rm a_{bin,PCE}}$ and ${\rm e_{bin,PCE}}$, and the modes of \aPCE and \ePCE from the end of the preceding, primary RLOF as initial conditions, we further explore the evolution of the CBP for the secondary RLOF and CE stage for the three cases where the binary evolves into a WD-WD pair with mass loss: i) \aCE = 5, NTCP; ii) \aCE = 10, TCP; iii) \aCE = 10, NTCP. The results are as follows.

During this secondary CE phase, for i) and ii) ${\rm \Psi(\aalpha)\approx0.4}$ and ${\rm \ac\approx2.4~\Msun / yr}$ -- comparable to \aalpha, indicating that the CBP evolves in the transition regime. For iii) ${\rm \Psi(\aalpha)\approx3.6}$, ${\rm \ac\approx0.3~\Msun / yr}$, and the evolution of the planet's orbit is in the runaway regime. The time of mass loss (${\rm T_{CE}}$) is close to the orbital period of the CBP, i.e. ${\rm T_{CE}/P_{CBP}\sim0.3-2.5}$. The binary shrinks by a factor of $\sim5-10$ into a WD-WD pair, and loses $\sim60-70\%$ of its mass. As a result, $\beta<\beta_{\rm eject}$ and the CBP should become unbound in a runaway regime. 

The evolution of \ap and \ep during the secondary CE phase is shown in Fig. \ref{fig:K1647sec}, where the upper panel represents case i) (\aCE = 5, NTCP), the middle panel case ii) (\aCE = 10, NTCP), and the lower panel -- case iii) (\aCE = 10, TCP). The CBP is ejected in more than $\sim80\%$ of the simulations. For \aalpha, the CBP's \ap expands to a mode value of $\sim8-15$ AU (depending on the scenario described above), with a maximum of up to 50-100 AU; the mode of \ep is in the range of (0.5-0.75), reaching a maximum of 0.95. For \aalphao, the CBP's orbit can expand up to $\sim30-50$ AU. 

By the end of the BSE simulations, the CBP can remain bound only for ii) as the other two scenarios trigger a SN shortly after the secondary CE. 

\begin{figure}
\centering
\epsscale{1.0}
\plotone{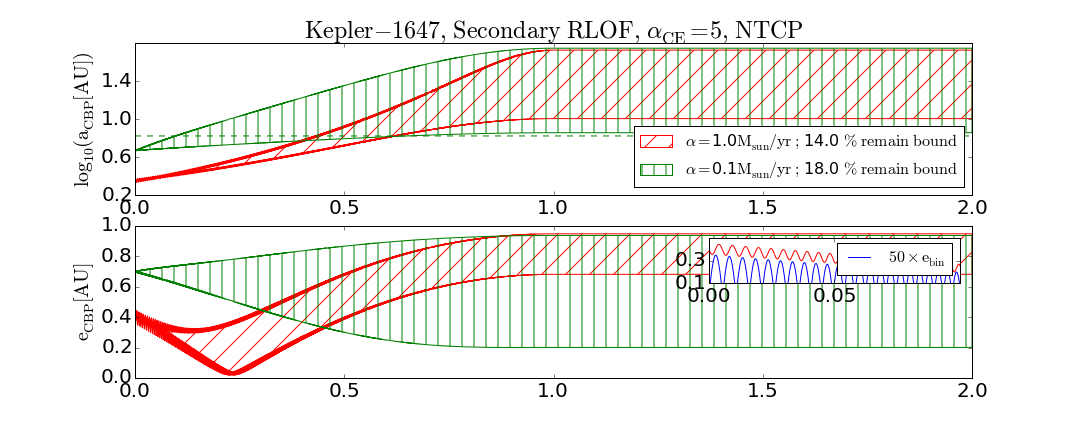}
\plotone{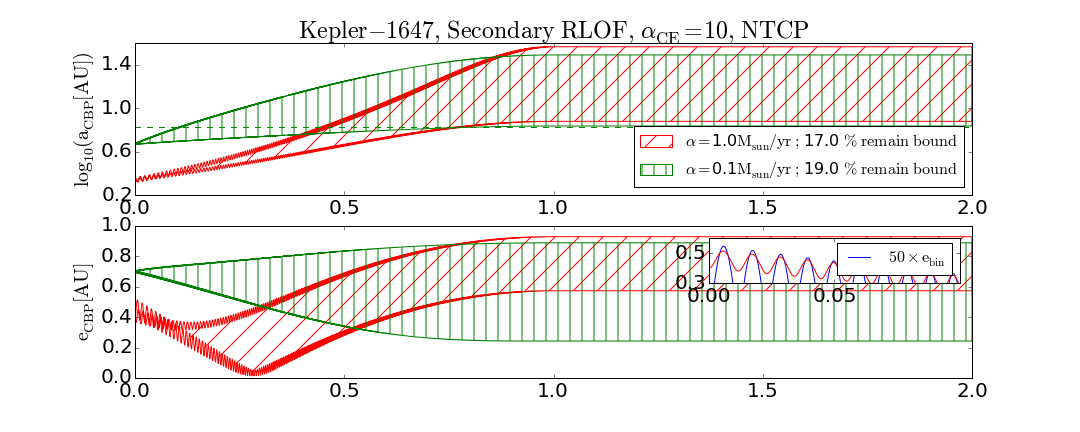}
\plotone{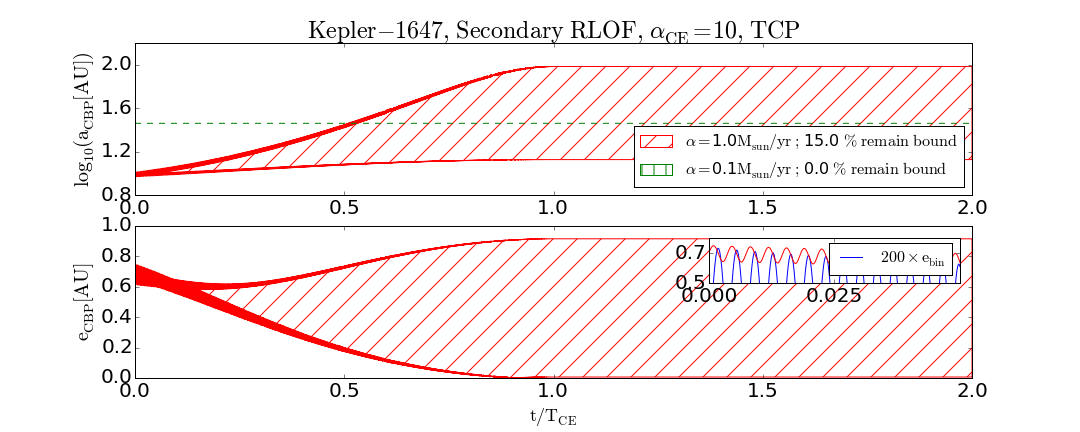}
\caption{Evolution of Kepler-1674 CBP's \ap and \ep during the secondary RLOF and CE for \aCE = 5, NTCP (upper panel), \aCE = 10, NTCP (middle panel) and \aCE = 10, TCP (lower panel) for \aalpha (red) and \aalphao (green). The planet is ejected in $\sim80-100\%$ of the simulations.
\label{fig:K1647sec}}
\end{figure}

\subsection{Kepler-35}

The system does not experience a CE evolution for {\bf $Z{nominal}$}, but the primary undergoes a RLOF and CE for ${\rm Z - 1\sigma_Z}$. Given the sensitivity of the binary evolution to the uncertainties in $Z$, here we outline the results for this system as well.

\subsubsection{The Binary}

The binary consists of 0.88\Msun + 0.81\Msun~stars on a slightly-eccentric ($e_{\rm bin,0}$ = 0.15), 21-day orbit which changes little in the first $\sim13$ Gyr. The CBP has an initial orbit of 0.6~AU. 

The primary star fills its Roche lobe at $t{\rm \approx12.99}$ Gyr and the binary enters a primary CE stage. During this stage $\Psi(\aalpha)\approx0.03$ and ${\rm \ac\approx29.6~\Msun / yr}$ -- much larger than \aalpha. As a result of the CE, the binary merges as a First Giant Branch star and loses $\sim35\%$ of its mass for \aCE = 0.5; here $\beta>\beta_{\rm eject}$, and the planet should remain bound in a runaway regime. The system does not experience significant mass loss by the end of the BSE simulations (15 Gyr), thus no major changes in \ap.

For \aCE=1/3/5/10 the binary orbit shrinks by a factor of $\sim 2-10$, loses $\sim35\%$ of its initial mass and evolves into a He-WD pair; here $\beta>\beta_{\rm eject}$, and the planet should remain bound in a runaway regime. 

\subsubsection{The CBP}

Overall, for \aalphao~the orbital evolution of the CBP is consistent with the adiabatic approximation. For \aalpha, mode(\aPCE)$\sim0.8-1$ AU and the planet gains an eccentricity of 0.2-0.4; the min/max ranges can be significant. In particular, for \aCE = 0.5, \aalpha,~and NTCP the CBP reaches sufficiently high eccentricities ($\ep > 0.95$) in $\sim4\%$ of the simulations to be ejected from the system; the CBP remains bound in all other cases. For \aalpha~and NTCP \aPCE can reach up to 7-10 AU, with $\ePCE > 0.9$.

\section{Discussion and Conclusions}
\label{sec:end}

Several key results emerge from our study:
 
i) Five of the nine \kepler CBP hosts experience at least one RLOF and a CE phase for their default parameters (masses, metallicities) by the end of our BSE simulations (15 Gyr); Kepler-35 experiences a primary RLOF for $Z - 1\sigma_Z$. Depending on the treatment of the CE stage, the binaries either merge after primary or secondary CE (typically for \aCE = 0.5/1) or evolve into very short-period WD-MS pairs (for \aCE = 1/3/5/10)\footnote{Which can also merge for a secondary-triggered CE phase.}; Kepler-1647 evolves into a WD-WD pair after a secondary RLOF and CE for \aCE = 5/10. Two systems trigger a double-degenerate Supernova explosion -- Kepler-34 for primary RLOF, and Kepler-1647 for a third RLOF and \aCE = 5, TCP, and \aCE = 10, NTCP. The binary systems lose a tremendous amount of mass -- up to $\sim60-70\%$ -- during the CE stages, and can shrink to sub-\Rsun orbital separations.

ii) From the dynamical perspective presented here, \kepler CBPs predominantly remain bound to their host binaries after the respective CE phases even for mass-loss rates as high as 1.0~${\rm \Msun~yr^{-1}}$. There are only 4 scenarios where the respective CBP has a 100\% probability of becoming unbound -- 1) for Kepler-34 where the SN explosion occurs with the first CE; 2) for the secondary CE phase of Kepler-1647 for \aalphao, \aCE = 10, TCP; 3) for the SN caused by the third RLOF of Kepler-1647, \aCE = 5, NTCP; and 4) SN caused by the third RLOF of Kepler-1647, \aCE = 10, TCP. In all other scenarios (106 total, not accounting for the initial phase differences $\Delta\theta_0$), the CBPs remain bound in the majority of cases (only Kepler-1647 has non-negligible ejection probability, again highly dependent on $\Delta\theta_0$). 

For mass-loss rates of 0.1~${\rm \Msun~yr^{-1}}$, the orbits of the CBPs evolve adiabatically -- well reproduced by REBOUNDx -- except for Kepler-1647. The orbital expansion of some of the \kepler CBPs is consistent with the adiabatic approximation even for mass-loss rates of 1.0~${\rm \Msun~yr^{-1}}$ (e.g. inner planet of Kepler-47, Kepler-64); in other cases, the mode of \aPCE is smaller than the adiabatic approximation (i.e. Kepler-1647). 

According to the analytical prescription, some systems should retain their CBPs even in a runaway mass-loss regime ($\Psi\gg1$). Our code fully reproduces the runaway approximation for single-star systems. 

iii) The transition regime ($\Psi\sim0.1-1$) -- where the evolution of some of \kepler CBPs falls for \aalpha -- is complex, and the final eccentricities and semi-major axes of the planets depend both on the treatment of the CE stage (i.e. \aCE, treatment of tides), and on the initial configuration of the system (i.e. $\Delta\theta_0$). We find that the orbits of \kepler CBPs can expand by more than an order of magnitude over the course of a single CBP year, reaching \aPCE of tens of AU (during the secondary CE of Kepler-1647 \ap can expand from $\sim10$ AU to $\sim100$ AU, see Table \ref{tab:temp_tab_CBP_fast_cont}). 

Multiplanet CBP systems add yet another level of complexity as they can experience both regimes simultaneously, e.g.~where the orbit of the inner Kepler-47 CBP expands adiabatically, the middle and outer CBP orbits grow non-adiabatically -- all during the primary CE phase. 

Overall, the CE-induced orbital evolution of CBPs is a dynamically rich and complex process in which a planet can migrate adiabatically during one CE phase, and non-adiabatically during another. As a result, a CBP cannot experience the same common envelope twice.

iv) If CE mass-loss rates are indeed high (e.g.~$\sim1.0~{\rm \Msun~yr^{-1}}$) -- as suggested by theoretical work -- and the orbits of \keplerp-like CBPs evolve (and survive) according to our simulations, we should expect to detect potentially highly eccentric planets orbiting PCE systems on very large orbits. Alternatively, if $\Psi\ll1$ then we should find PCE CBPs on low-eccentricity orbits. Interestingly, recent observational efforts have produced a number of PCE Eclipse-Time Variations (ETV) CBP candidates on wide, notably eccentric orbits (e.g.,~\cite{zorotovic13} and references therein), suggestion a possible connection with PCE \keplerp-like CBPs.

On Fig. \ref{fig:Kepler_vs_PCECBPs} we show \aPCE vs \ePCE of {\it Kepler's} CBPs\footnote{For Kepler-47 we only show the middle planet (Planet 2) as it has the highest probability to survive both the CE phase and subsequent planet-planet interactions.} and compare them to those of the PCE ETV CBP candidates\footnote{For those candidates that do not have published eccentricities we set the respective \ep to zero.}. We caution that the  comparison is not direct as neither the CE stages of the different \kepler systems occur all at the same time, nor do the PCE ETV CBPs have identical ages. Instead, given the handful of known targets -- {\it Kepler's} nine compared to estimated millions of similar CBPs \citep{welsh12}; a dozen PCE ETV CBPs -- the lower two panels on the figure portray a potential distribution of an underlying PCE CBP population with a range of ages and evolutionary stages. 

Overall, the PCE orbital configurations of {\it Kepler's} CBPs are qualitatively consistent with those of the observed population of the currently known PCE ETV CBP candidates\footnote{With the caveat that the latter are much too massive.}. Thus our results both assist in interpreting the nature of these candidates and in guiding future observational efforts to discover new systems. Such discoveries will provide a deeper understanding of the evolution of planets in binary star system, and also much-needed observational constraints on the stellar astrophysics of the complex CE phase, as {\it``CE is one of the most important unsolved problems in stellar evolution''}, according to \cite{ivanova13},{\it``and is arguably the most significant and least-well-constrained major process in binary evolution''} (but also see \cite{taam00, taam10} and\cite{webbink08} for alternative reviews). 

\begin{figure}
\centering
\epsscale{1.1}
\plotone{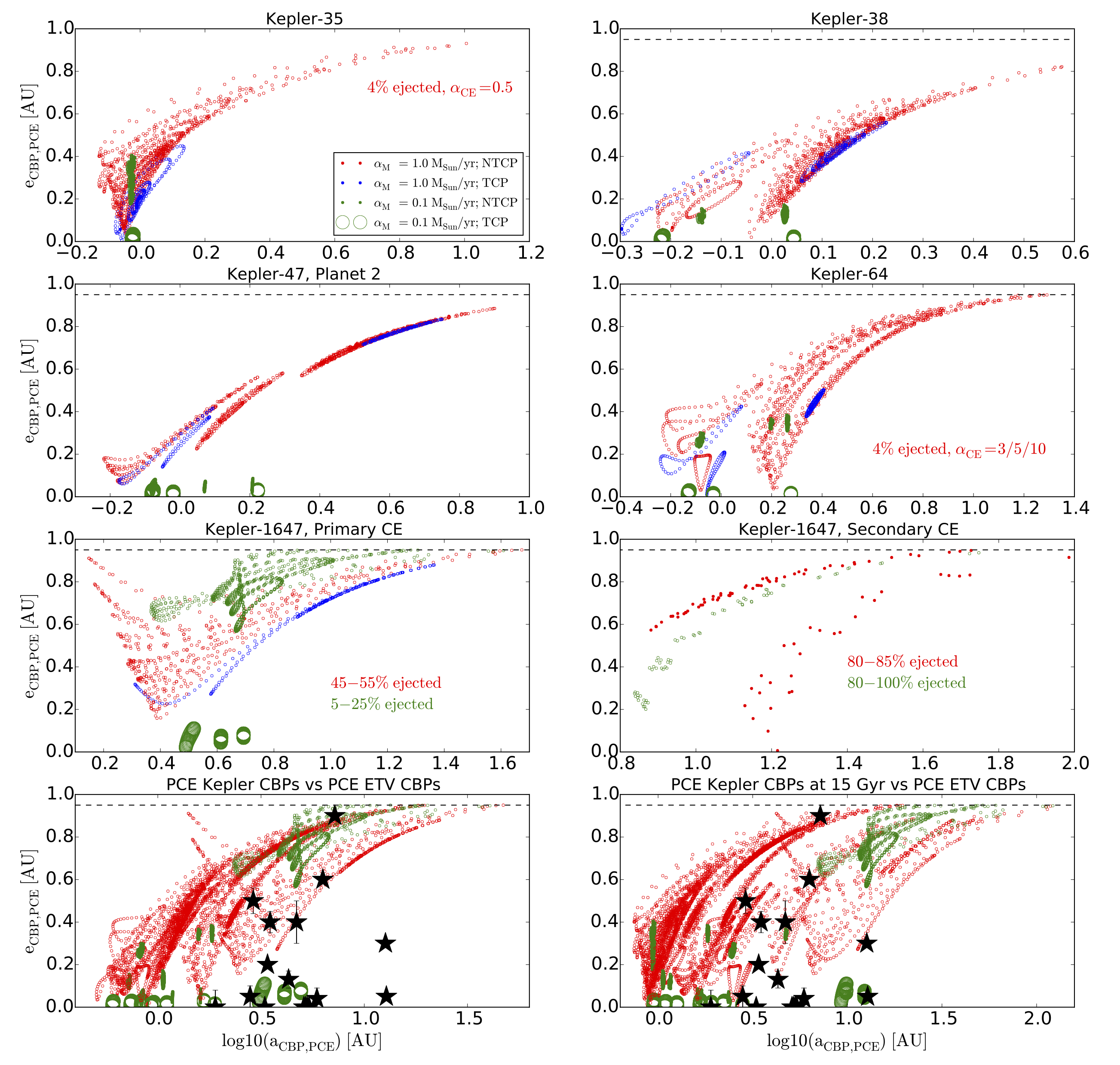}
\caption{Upper six panels: \aPCE vs \ePCE for {\it Kepler's} CBPs for various CE treatments (green, blue and red symbols) and initial conditions (${\rm \Delta\theta_0=0\div1}$), along with the respective ejection probabilities; lower two panels: comparison between {\it Kepler's} \aPCE vs \ePCE and those of the currently known PCE ETV CBPs candidates (black symbols) at the ends of the respective \kepler CEs (left) and at the end of our BSE simulations (15 Gyr, right). Given the associated observational and numerical uncertainties, our forward-evolution results are qualitatively consistent with the observed \aPCE vs \ePCE distributions. See text for details.
\label{fig:Kepler_vs_PCECBPs}}
\end{figure}

vi) A CB body gravitationally perturbs its host binary star, and if the latter is eclipsing these perturbations can be manifested as deviations from linear ephemeris in the measured stellar eclipse times. The cause of these variations can be either dynamical, where the third body's influence changes the orbital elements of the binary star, or a light travel-time effect (LTTE), where the tertiary object and the binary revolve around a common center of mass. ETVs are indeed a powerful -- and highly productive -- method to discover and study stellar triple and higher order systems (e.g.~\citealt{orosz15, borkovits15}, and references therein), and have recently been used to study CBPs as well. In particular, the former effect has been measured for several of the \kepler CBP systems and used to constrain the respective planetary masses \citep{doyle11, welsh12, kostov16}. The latter effect has been suggested as the cause for measured ETVs for a number of PCE EB systems with proposed CB companions (as mentioned above, also see~\cite{volschow16} and references therein). 

Assuming that the orbits of the \kepler CBPs studied here evolve only dynamically and their masses remain constant (i.e. no changes due to, for example, interactions with a CE-triggered CB disk), it is informative to evaluate the respective planets' ETV-based detectability after the CE stages of the relevant systems. To calculate the expected amplitudes of the two effects discussed above, $A_{\rm dyn}$ and $A_{\rm LTTE}$, we use the formalism of \cite{borkovits12, borkovits15}. 

The respective amplitudes and periods for Kepler-35 (\Mplan = 0.13\Mjup) and Kepler-1647 (\Mplan = 1.5\Mjup) -- the only two systems with well-constrained CBP masses -- are listed in Table \ref{tab:etvs} (for the respective range of binary and CBP parameters listed in Tables \ref{tab:temp_tab_CBP_fast_cont}), and compared to the detected signals for the two PCE CBP candidates NN Ser and HW Vir. As seen from Table \ref{tab:etvs}, detection of such ETVs is feasible both in terms of the amplitudes and periods of the expected signal, which are qualitatively similar to the ETV signals of the two PCE CBP candidates. 

Another option for the detection of a PCE CBP is to directly image the planet. The benefits for such detection are two fold: a) the contrast ratio between the binary and the CBP decreases after each CE phase as the primary and/or secondary evolves from a luminous MS star to a much fainter WD; b) the angular separation between the binary and the CBP increases\footnote{Such that the minimum increase is typically for adiabatic orbital expansion.} (e.g.~\citealt{hardy15}). There is potentially a third benefit, where the CBP can accrete mass after the CE stage (e.g.~\citealt{zorotovic13, bear14}) and becomes brighter, thus further decreasing the contrast ratio. 

Recently, \cite{hardy15} used VLT/SPHERE to observe the V471 Tau system -- an eclipsing binary star with measured ETVs -- with the goal to directly image a sub-stellar mass CB candidate suspected to be the cause of the ETVs. Given the known age of the system, and the masses of the binary and the CB candidate, direct detection of such tertiary body should have been well within the capabilities of the instrument. \cite{hardy15}, however, report a null detection, casting doubt on the interpretation of the measured ETV signal (but see \cite{vaccaro15} for alternative explanation). Nevertheless, based on our results for the dynamical survivability of CBPs around evolving binary stars, and on estimates of the occurrence rates of planets orbiting short-period MS binary stars (e.g.~\citealt{welsh12}), we encourage the continuation of such direct imaging efforts -- aimed at both PCE, very short-period binaries, and also at PCE WD-coalesced binaries where the contrast ratio is even more favorable. 

vii) The closer a planet's orbit is to a single star during the MS stage, the worst its prospects are for survival when the star ascends the RGB and AGB. For example, Mercury, with an orbit of 0.46~AU will be destroyed as the Sun evolves off the main sequence (e.g.~\citealt{schroder08}. In contrast, our results suggest that despite their dramatic evolution, close binary stars may in fact be better for the survival of planets compared to single stars. For example, the Kepler-38 CBP has an initial orbital separation similar to Mercury's, and yet it survives the evolution of a binary star with a total mass of $\sim1~\Msun$. A CBP with an even smaller orbit -- as small as 0.3-0.35~AU -- can survive the evolution of 1.3-1.4~\Msun~(total mass) central binary star as is the case for the inner planet in Kepler-47. A planet with the same orbit around a single star of the same mass will have no chance for survival \citep{villaver07}. Thus despite being a disruptive stage for the binary star itself, the CE phase may in fact promote planet survivability. 

viii) With the exception of Kepler-1647, the \kepler CBPs currently orbit their binary star hosts within a factor of 2 of the critical limit for dynamical stability (\cite{holman99}, HW hereafter), i.e. $a_{\rm CBP,0}(1-e_{\rm CBP, 0})/a_{crit, 0}\sim2$ (the factor for Kepler-1647 is $\sim7$). However, as the binaries shrink and lose mass during the CE, and the planets migrate to larger, eccentric orbits, this ratio will change. To calculate ${\rm a_{crit, PCE}}$ -- the PCE critical limits (for those scenarios where the binaries do not coalesce, i.e. \aCE = 3/5/10) --  we use Eqn. 3 from HW using a binary mass ratio ${\rm \mu = M_{A,PCE} / (M_{A,PCE} + M_{B,PCE})}$\footnote{Since the primary star becomes the less massive component of the binary after the CE phase.}. As the CE phases result in a variety of \abin, \aPCE and \ePCE, here we quote only the most constraining critical limits, i.e. where the respective separation between \abin and ${\rm \aPCE(1-\ePCE)}$ is at a minimum. For \aalpha, the limits for Kepler-38, Kepler-47, Kepler-64, Kepler-1647 (NTCP) and Kepler-1647 (TCP) in terms of ${\rm \aPCE(1-\ePCE)/a_{crit, PCE}}$ are 3.3, 7.9, 5.3, 15.9, 14.7 and 29.5 respectively, indicating that the CBPs remain in a dynamically-stable regime. 

While it is beyond the scope of this study, we note that tidal evolution of the binary orbit prior to the CE phase may affect the CBP orbits. Specifically, as most of the planets are currently orbiting close to their host binaries, a pre-CE decay in \abin means that strong mean motion resonances (MMR) may sweep over the planet, leading to eccentricity excitation and potential destabilization. As an example, \abin of Kepler-38 decreases from 31.6~\Rsun to 24.6~\Rsun prior to the CE phase for TCP (see Table \ref{tab:BSE_example}), indicating that the CBP's orbit crosses 6:1, 7:1 and 8:1 MMR. The MMR crossings are a) 7:1, 8:1, 9:1 for Kepler-47 (Planet 1); b) 7:1 through 11:1 for Kepler-64; and c) 99:1 through 116:1 for Kepler-1647. 

ix) {\it ``...It is interesting to consider the situation at a much earlier time in the past when the primary was near the zero-age main sequence...}'', note \cite{orosz12a} for the case of Kepler-38b, {\it ``The primary's luminosity would have a factor $\approx3$ smaller...''}. While Kepler-38b is currently closer to its binary host than the inner edge of the habitable zone (HZ) for the system, four of the \kepler CBPs (Kepler-16, Kepler-47c, Kepler-453, Kepler-1647) reside in the HZ. As their host binaries evolve, however, the location of the HZ will change. 

Unlike the case for single stars where both the HZ and a (surviving) planet's orbit will expand during the RGB and AGB stages \citep{villaver07}, as a close binary star evolves through a CE stage the HZ can shrink while its CBP migrates outwards. The shrinking is caused by the luminosity drop of the central binary as one of its stars rapidly transitions from the MS to a compact object. Thus a CBP residing in the HZ during the pre-CE binary will leave the zone after the CE phase (e.g.~Kepler-1647), whereas a CBP that is initially interior to the HZ may migrate to a PCE orbit coinciding with the PCE HZ. 

Consequently, it would be equally interesting to consider the configuration of a CBP system at a much later time in the binary evolution. For example, the current orbital configuration of Kepler-35b is such that the planet's orbit is internal to the HZ \citep{welsh12}. However, as its host binary evolves through the CE phase, for \aCE = 1/3/5/10 the primary star evolves into a HeWD (while the secondary remains on the MS as a 0.81~\Msun~star) and the CBP migrates to ${\rm mode(\aPCE)=0.84-1.0~AU, mode(\ePCE)=0.0-0.4}$ (see Tables \ref{tab:temp_tab_CBP_fast_cont}). This migration places the planet near the conservative HZ of ${\rm 0.98-1.77}$ AU (for 1 ${\rm M_\Earth}$ \citep{kopparapu14}, using the luminosity, ${\rm 0.4~L_{Sun}}$, and temperature of the secondary star, 5200 K \citep{welsh12}, and ignoring the flux contribution of the WD).

\subsection{Limitations}
\label{sec:limitations}

The results presented here are based on the assumptions that the CBPs do not interact with the material ejected from their binary star. However, numerical studies have indicated that this material is neither lost isotropically from the binary during the CE phase, nor does it all become unbound. Instead, $1-10\%$ of the ejecta may fall back into a CB disk according to \cite{kashi11}, and \cite{passy12} suggest that $\sim80\%$ of the ejected material may remain gravitationally bound to the binary (also see \citealt{kuiper41, shu79}, and \citealt{pejcha16}   for mass loss outflows through the ${\rm L_2}$ Lagrange point\footnote{Mass-loss through ${\rm L_2}$ results in several possible outcomes, e.g. isotropic or equatorial wind, CB disk -- for details see Tab. 1, Fig. 12 and 13 of \citealt{pejcha16}.}). Either of these scenarios would significantly complicate the dynamical evolution of the system as the CBP could accrete material and gain mass, and also experience migration similar to that during planetary formation\footnote{There are, however, two potential benefits of the former in terms of detection: a more luminous planet would be more amenable to direct imaging efforts; and a more massive planet would cause stronger ETVs.}. Such accretion of material of a different specific angular momentum will change the orbital evolution of the planet, as will gravitational interaction with the bulk gas in an accretion-favorable environment such as a CB disc. Additionally, interactions between CBPs and the CE ejecta of close binary stars (including ejection or destruction of the planets) may also play an important role in the elusive mechanisms responsible for the shaping, morphology and chemistry of planetary nebulae (see \cite{bear16} for triple star origin of asymmetric planetary nebulae; for recent reviews see \cite{zijlstra14} and \cite{jones15}, and references therein). We note that the CE-triggered dynamical disturbances discussed here occur on timescales of $\sim1$~year, and are thus `instantaneous' compared to any subsequent planet migration, which occurs on much longer timescales. 

PCE CB disks are also suspected to trigger a new episode of planet formation, i.e.~a second generation scenario (e.g.~\citealt{verhoelst07, perets10, tutukov12} and references therein), and have also been suggested to be the birthplace of several PCE CBP candidates \citep{volschow13, schleicher14}\footnote{Second generation has also been proposed for the origin of the first detected exoplanet \citep{wolszczan92}.}. Indeed, a recent discovery of dust around NN Ser supports the presence of a remnant PCE disc \citep{hardy16}. Alternatively, based on angular momentum constraints \cite{bear14} argue that these CBP candidates formed together with their host binary stars (i.e.~first generation). Mixed scenarios, where the CBP forms with the binary, survives the CE stage and subsequently acts as a seed for the accretion of more mass, are also possible \citep{armitage99, perets10, schleicher15}, showcasing the complexity and diversity of the formation and evolution mechanisms of planets around close binary stars. 

These complications are, however, beyond the scope of our study and we defer their implementation to future work.

\subsection{Conclusions}

We have presented numerical studies of the dynamical evolution of the nine \kepler CBP systems as their host binary stars undergo CE phases. Five of the systems undergo at least one RLOF and CE phases; Kepler-1647 experiences three RLOFs. Two systems trigger a double-degenerate SN explosion. Depending on the treatment of the CE phase (i.e. efficiency and tidal evolution), the binaries either coalesce or shrink into very close WD-MS or WD-WD pairs of stars. Despite the dramatic reconfiguration of the binary stars during these violent evolutionary phases, the planets predominantly survive the CE stage even for mass-loss rates of ${\rm 1~\Msun~yr^{-1}}$. The PCE orbital configurations of these CBPs depend on the rate at which the binary loses mass. For mass-loss rates of ${\rm 1~\Msun~yr^{-1}}$ the CBPs migrate non-adiabatically outwards to highly eccentric orbits, while for slower mass-loss rates (${\rm 0.1~\Msun~yr^{-1}}$) our simulations reproduce the expected adiabatic orbital expansion. Overall, the PCE semi-major axes and eccentricities of \kepler CBPs are qualitatively consistent with those of the currently known PCE CBP candidates. The mode of orbital expansion depends on the particular configuration of the system, such that a CBP can migrate adiabatically during the primary CE stage, but non-adiabatically during the secondary CE stage. Multiplanet CBP systems can experience both modes simultaneously, i.e. an inner planet can migrate adiabatically while an outer does so non-adiabatically -- both occurring during the same CE stage. Our results also indicate that planets are more likely to survive around evolving close binary stars than around evolving single stars, thus improving the discovery prospects for CBPs in PCE systems. 

\acknowledgments

We thank the anonymous referee for the insightful comments that helped us improve this paper. The authors are grateful to Jarrod Hurley and Marten van Kerkwijk for valuable discussions. VBK gratefully acknowledges support by an appointment to the NASA Postdoctoral Program at the Goddard Space Flight Center. DT was supported by a postdoctoral fellowship from the Centre for Planetary Sciences at the University of Toronto at Scarborough, and is grateful for additional support from the Jeffrey L. Bishop Fellowship. This work was supported in part by NSERC grants to RJ. We acknowledge conversations with Daniel Fabrycky, Nader Haghighipour, Kaitlin Kratter, Boyana Lilian, Jerome Orosz, William Welsh.

%

\bibliographystyle{apj}

\newpage
\appendix
\section{\\List of Abbreviations and Parameters} \label{App:AppendixA}
{\noindent BSE: Binary Star Evolution code\\}
CE: Common Envelope \\
CBP: CircumBinary Planet\\
EB: Eclipsing Binary\\
NTCP: No Tidal Circularization Path (tides ``OFF'' in BSE)\\
PCE: Post-Common Envelope\\
RLOF: Roche Lobe Overflow \\
TCP: Tidal Circularization Path (tides ``ON'' in BSE)\\
\\
{\noindent \aM: Common Envelope mass-loss rate \\}
\aalpha: 1~\Msun/yr mass-loss rate \\
\aalphao: 0.1~\Msun/yr mass-loss rate \\
${\rm \alpha_{crit}}$: Critical Common Envelope mass-loss rate \\
\aCE: Common Envelope efficiency parameter \\
${\rm a_{CBP,0}}$: Initial semi-major axis of the circumbinary planets \\
${\rm a_{CBP,PCE}}$: Semi-major axis of the circumbinary planets at the end of the Common Envelope\\
${\rm \beta}$: Ratio between initial and final mass of the system \\
${\rm \beta_{eject}}$: Runaway ejection ratio between initial and final mass of the system \\
${\rm e_{CBP,0}}$: Initial eccentricity of the circumbinary planets \\
${\rm e_{CBP,PCE}}$: Eccentricity of the circumbinary planets at the end of the Common Envelope phase\\
${\rm \psi}$: Common Envelope mass-loss index \\
\tCE: Common Envelope mass-loss timescale \\

\newpage

\begin{table}[ht]
\begin{center}
\footnotesize
\caption{BSE Parameter Space
\label{tab:bseparams}}
\begin{tabular}{ccc}
\hline
\hline
Parameter & Explored Values \\
\hline
\hline
$\Mprim$ & $\mathbf{M_1}, \Mprim + 3\sigma, \Mprim - 3\sigma$ \\
$\Msec$ & $\mathbf{M_2}, \Msec + 3\sigma, \Msec - 3\sigma$ \\
Metallicity $Z$ & $\mathbf{Z}, Z + 1\sigma, Z + 2\sigma, Z + 3\sigma, Z - 1\sigma, Z - 2\sigma, Z - 3\sigma$ \\
$\alpha_{CE}$ & $0.5, 1.0, \mathbf{3.0}, 5.0, 10.0$ \\
Tides & OFF, \textbf{ON} \\
de Kool CE Model & \textbf{OFF}, ON \\
Force Corotation & \textbf{OFF}, ON \\

\hline
\multicolumn{3}{l}{Note: Values denoted in boldface represent the default parameters used in our simulations.} \\
\hline
\end{tabular}
\\
\end{center}
\end{table}

\begin{table}[ht]
\begin{center}
\caption{BSE Stellar Types
\label{tab:bse_stellar_types}}
\begin{tabular}{cc}
\hline
\hline
Stellar Type & Description \\
\hline
\hline
0 & Deeply or fully convective low-mass MS star \\
1 & Main Sequence star \\
2 & Hertzsprung Gap (HG) \\
3 & First Giant Branch (GB) \\
4 & Core Helium Burning (CHeB) \\
5 & First/Early Asymptotic Giant Branch (EAGB) \\
6 & Second/Thermally Pulsing Asymptotic Giant Branch (TPAGB) \\
7 & Main Sequence Naked Helium star (HeMS) \\
8 & Hertzsprung Gap Naked Helium star (HeHG) \\
9 & Giant Branch Naked Helium Star (HeGB) \\
10 & Helium White Dwarf (HeWD) \\
11 & Carbon/Oxygen White Dwarf (COWD) \\
12 & Oxygen/Neon White Dwarf (ONeWD) \\
13 & Neutron Star (NS) \\
14 & Black Hole (BH) \\
15 & Massless Supernova/Massless Remnant \\

\hline
\end{tabular}
\\
\end{center}
\end{table}

\begin{table}[ht]
\begin{center}
\scriptsize
\caption{BSE Output Evolution Stages
\label{tab:bseoes}}
\begin{tabular}{cc}
\hline
\hline
Evolutionary Stage Label & Description \\
\hline
\hline
INITIAL & Initial configuration \\
KW CHNGE & Stellar type change \\
BEG RCHE & Begin Roche lobe overflow \\
END RCHE & End Roche lobe overflow \\
CONTACT & Contact system \\
COELESCE & Coalescence of stars \\
COMENV & Common-envelope system \\
GNTAGE & New giant star from CE; appropriate age and initial mass set to match core-mass and stellar mass \\
NO REMNT & No remnant \\
MAX TIME & Max evolutionary time reached; end of program \\
DISRUPT & System is disrupted \\
BEG SYMB & Begin symbiotic system \\
END SYMB & End symbiotic system \\
BEG BSS & Begin blue stragglers \\
\hline
\end{tabular}
\\
\end{center}
\end{table}

\clearpage
\begin{table}[ht]
\begin{center}
\scriptsize
\caption{BSE evolution of Kepler-38 for \aCE = 0.5. The BSE results for our entire set of simulations for all \kepler systems are presented as an online supplement.
\label{tab:BSE_example}}
\begin{tabular}{cccccccccc}
\hline
\hline
Time$\dagger\dagger$ & \Mprim & \Msec & Stell. Type & Stell. Type & \abin & \ebin & ${\rm R_1/R_{Roche}}$ & ${\rm R_1/R_{Roche}}$ & Evol. Stage \\
${\rm [Gyr]}$ & [\Msun] &  [\Msun] &  &  & [\Rsun] & & & & \\
\hline
Kepler-38$\dagger$ & & & & & & & & \\
0.0 & 0.95 & 0.25 & 1 & 0 & 31.6 & 0.10 & 0.05 & 0.03 & INITIAL \\ 
12.36 & 0.95 & 0.25 & 2 & 0 & 31.6 & 0.10 & 0.10 & 0.03 & KW CHNGE \\
13.01 & 0.95 & 0.25 & 3 & 0 & 31.61 & 0.10 & 0.14 & 0.03 & KW CHNGE \\
13.72 & 0.94 & 0.25 & 3 & 0 & 31.86 & 0.10 & 1.00 & 0.03&BEG RCHE \\
13.72 & 0.75 & 0.25 & 3 & 15 & 0.51 & 0.00 & 1.00 & 0.03 & COMENV \\
13.76 & 0.47 & 0.00 & 10 & 15 & 0.00 & -1.00 & 0.00 & -1.00 & KW CHNGE \\
15.00 & 0.47 & 0.00 & 10 & 15 & 0.00 & -1.00 & 0.00 & -1.00 & MAX TIME \\
\hline
Kepler-38$\ddagger$ & & & & & & & & \\
0.0 & 0.95 & 0.25 & 1 & 0 & 31.59 & 0.10 & 0.05 & 0.03 & INITIAL \\ 
12.36 & 0.95 & 0.25 & 2 & 0 & 31.60 & 0.10 & 0.10 & 0.03 & KW CHNGE \\ 
13.01 & 0.95 & 0.25 & 3 & 0 & 31.45 & 0.09 & 0.14 & 0.03 & KW CHNGE \\ 
13.70 & 0.94 & 0.25 & 3 & 0 & 24.58 & 0.00 & 1.00 & 0.04 & BEG RCHE \\ 
13.70 & 0.91 & 0.25 & 3 & 15 & 0.37 & 0.00 & 1.00 & 0.04 & COMENV   \\ 
13.75 & 0.66 & 0.00 & 4 & 15 & 0.00 & -1.00 & 0.00 & -1.00 & KW CHNGE \\ 
13.88 & 0.63 & 0.00 & 5 & 15 & 0.00 & -1.00 & 0.00 & -1.00 & KW CHNGE \\ 
13.89 & 0.55 & 0.00 & 6 & 15 & 0.00 & -1.00 & 0.00 & -1.00 & KW CHNGE \\ 
13.89 & 0.52 & 0.00 & 11 & 15 & 0.00 & -1.00 & 0.00 & -1.00 & KW CHNGE \\ 
15.00 & 0.52 & 0.00 & 11 & 15 & 0.00 & -1.00 & 0.00 & -1.00 & MAX TIME \\ 
\hline
\hline
\multicolumn{10}{l}{$\dagger\dagger$: Time in the online supplement is in Myr.} \\
\multicolumn{10}{l}{$\dagger$: NTCP.} \\
\multicolumn{10}{l}{$\ddagger$: TCP.} \\
\end{tabular}
\\
\end{center}
\end{table}

\begin{table}[ht]
\begin{center}
\caption{Kepler System Default Initial Parameters.
\label{tab:keplerinitial}}
\begin{tabular}{c|cccccccc}
\hline
\hline
Name & Kepler ID & \Mprim & \Msec & ${\rm P_{bin}}$ & $\ebin$ & $Z$ & ${\rm P_{CBP}}$ & $\ep$  \\
 & & [\Msun] & [\Msun] & [days] & & & [days] & \\
\hline
\hline
Kepler-16 & 12644769 & 0.69 & 0.20 & 41.08 & 0.16 & 0.010 & 228.78 & 0.01 \\ 
Kepler-34 & 8572936 & 1.05 & 1.0208 & 27.79 & 0.52 & 0.016 & 288.82 & 0.18 \\
Kepler-35 & 9837578 & 0.89 & 0.81 & 20.73 & 0.14 & 0.009 & 131.46 & 0.04 \\
Kepler-38 & 6762829 & 0.95 & 0.25 & 18.79 & 0.1 & 0.015 & 105.59 & 0.03 \\
Kepler-47$\ddagger$ & 10020423 & 0.96 & 0.34 & 7.45 & 0.02 & 0.011 & 49.5 & 0.03 (P1) \\
& & & & & & & 187.4 & 0.02 (P2) \\
& & & & & & & 303.16 & 0.04 (P3) \\
Kepler-64 & 4862625 & 1.53 & 0.41 & 20 & 0.22 & 0.031$\dagger\dagger$ & 138.5 & 0.1 \\
Kepler-413 & 12351927 & 0.82 & 0.54 & 10.12 & 0.04 & 0.012 & 66.26 & 0.12 \\
Kepler-453 & 9632895 & 0.94 & 0.2 & 27.34 & 0.05 & 0.023 & 240.5 & 0.04 \\
Kepler-1647 & 5473556 & 1.22 & 0.97 & 11.259 & 0.15 & 0.014 & 1107.59 & 0.06 \\
\hline
\hline
\multicolumn{9}{l}{$\ddagger$: J. Orosz, priv. communication.} \\
\multicolumn{9}{l}{$\dagger\dagger$: BSE allows maximum metallicity of $Z = 0.03$.} \\
\end{tabular}
\\
\end{center}
\end{table}

\clearpage
\begin{table}[ht]
\begin{center}
\scriptsize
\caption{Binary Evolution from BSE. The numbers quoted in parenthesis for the time represent the corresponding range, obtained from BSE for ${\rm Z\pm3\sigma_Z}$ (see text for details). The typical range for ${\rm \Mprim and~\Msec}$ is ${\rm \sim 0.01-0.1~\Msun}$, and ${\rm \sim 0.1-1~\Rsun}$ for ${\rm a_{bin}}$.
\label{tab:temp_tab_EB}}
\begin{tabular}{cc||ccccc||c}
\hline
\hline
${\rm \alpha_{CE}}$ & Tides & ${\rm \Mprim}$ & ${\rm \Msec}$ & \abin & \ebin & Time RLOF$^{\S}$ & Notes \\
 &  & [${\rm M_{Sun}}$] & [${\rm M_{Sun}}$] & [\Rsun] & & [Gyr] &  \\
\hline
\hline
Kepler-34 & & & & & & & \\
 -- & -- & 1.05 & 1.02 & 49.2 & 0.52 & 0 & \\
0.5/1/3/5/10 & NTCP & -- & -- & -- & -- & 9.64(1.7) & DD SN \\
0.5/1/3/5/10 & TCP & -- & -- & -- & -- & 9.66(1.7) & DD SN$^\dagger$ \\
\hline
Kepler-38 & & & & & & & \\
 -- & -- & 0.95 & 0.25 & 31.6 & 0.1 & 0 & \\
0.5 & NTCP & 0.76 & -- & -- & -- & 13.72(0.6) & CE + Merger \\
0.5 & TCP & 0.91 & -- & -- & -- & 13.70(0.6) & CE + Merger \\
1/3/5/10 & NTCP & 0.27 & 0.25 & 1.0/2.8/4.5/7.8 & -- & 13.72(1.2) & CE \\
1/3/5/10 & TCP & 0.25 & 0.25 & 0.7/2.0/3.2/5.7 & -- & 13.70(1.1) & CE$^{\S\S}$ \\
\hline
Kepler-47 & & & & & & & \\
 -- & -- & 0.96 & 0.34 & 17.5 & 0.02 & 0 &  \\
0.5 & NTCP & 1.08 & -- & -- & -- & 12.03(1.3) & CE + Merger \\
0.5 & TCP & 1.14 & -- & -- & -- & 11.99(1.3) & CE + Merger \\
1.0 & NTCP & 0.78 & -- & -- & -- & 12.03(1.5) & CE + Merger \\
1.0 & TCP & 0.96 & -- & -- & -- & 11.99(1.5) & CE + Merger \\
3/5/10 & NTCP & 0.23 & 0.34 & 1.7/2.6/4.5 & -- & 12.03(1.7) & CE \\
3/5/10 & TCP & 0.21 & 0.34 & 1.2/1.9/3.4 & -- & 11.99(1.6) & CE \\
\hline
Kepler-64 & & & & & & & \\
 -- & -- & 1.53 & 0.41 & 38.7 & 0.22 & 0 &  \\
0.5 & NTCP & 1.53 & -- & -- & -- & 3.02(0.3) & CE + Merger \\
0.5 & TCP & 1.68 & -- & -- & -- & 3.02(0.3)$^{\dagger\dagger}$ & CE + Merger \\
1.0 & NTCP & 0.8 & -- & -- & -- & 3.02(0.3) & CE + Merger \\
1.0 & TCP & 1.35 & -- & -- & -- & 3.02(0.3) & CE + Merger \\
3/5/10 & NTCP & 0.28 & 0.41 & 2.4/3.9/7.0 & -- & 3.02(0.3) & CE \\
3/5/10 & TCP & 0.26 & 0.41 & 1.6/2.6/4.7 & -- & 3.02(0.3) & CE \\
3 & TCP & 0.67 & -- & -- & -- & 6.20(0.4) & CE$^\ddagger$ + Merger \\
\hline
Kepler-1647 & & & & & & & \\
 -- & -- & 1.22 & 0.97 & 27.4 & 0.15 & 0 &  \\
0.5 & NTCP & 1.81 & -- & -- & -- & 5.45(0.45) & CE + Merger \\
0.5 & TCP & 1.87 & -- & -- & -- & 5.44(0.46) & CE + Merger \\
1.0 & NTCP & 1.28 & -- & -- & -- & 5.45(0.44) & CE + Merger \\
1.0 & TCP & 1.45 & -- & -- & -- & 5.44(0.43) & CE + Merger \\
3/5/10 & NTCP & 0.25 & 0.97 & 4.0/6.2/10.1 & -- & 5.45(0.45) & CE \\
3/5/10 & TCP & 0.24 & 0.97 & 3.6/5.4/8.9 & -- & 5.44(0.36) & CE \\
3 & TCP & 1.16 & -- & -- & -- & 6.07(0.35) & CE$^\ddagger$ + Merger \\
5 & NTCP & 0.25 & 0.17 & 0.6 & -- & 12.06(0.9) & CE$^\ddagger$ \\
5 & NTCP & -- & -- & -- & -- & 12.93(1.0) & DD SN \\
5 & TCP & 1.15 & -- & -- & -- & 9.85(0.5) & CE$^\ddagger$ + Merger \\
10 & NTCP & 0.25 & 0.2 & 1.9 & -- & 12.27(1.1) & CE$^\ddagger$ \\
10 & TCP & 0.24 & 0.16 & 0.7 & -- & 11.86(0.9) & CE$^\ddagger$ \\
10 & TCP & -- & -- & -- & -- & 13.99(0.5) & DD SN \\
\hline
Kepler-35 & & & & & & & (${\rm Z - 1\sigma_Z}$) \\
 -- & -- & 0.88 & 0.81 & 37.9 & 0.14 & 0 & \\
0.5 & NTCP & 1.1 & -- & -- & -- & 12.99(1.44) & CE + Merger \\
0.5 & TCP & 1.09 & -- & -- & -- & 12.99(1.44) & CE + Merger \\
1/3/5/10 & NTCP & 0.29 & 0.81 & 3.5/8.8/12.7/19.1 & -- & 12.99(1.44) & CE \\
1/3/5/10 & TCP & 0.28 & 0.81 & 3.1/7.9/11.5/17.4 & -- & 12.99(1.44) & CE \\
\hline
\hline
\multicolumn{8}{l}{$^{\S}$: Roche Lobe OverFlow.} \\
\multicolumn{8}{l}{$^\dagger$: Double-degenerate Supernova.} \\
\multicolumn{8}{l}{$^\ddagger$: Secondary CE.} \\
\multicolumn{8}{l}{$^{\S\S}$: Second Roche Lobe fill and merger shortly after first CE for $\alpha_{CE}=1$.} \\
\multicolumn{8}{l}{$^{\dagger\dagger}$: Only lower range because BSE does not accept Z $>$ 0.3 (the metallicity of Kepler-64).} \\
\end{tabular}
\\
\end{center}
\end{table}

\clearpage
{\renewcommand{\arraystretch}{1.55}%
\begin{table}[ht]
\begin{center}
\scriptsize
\caption{PCE CBP's semi-major axis and eccentricity for ${\rm \aM = 1.0~M_{Sun}/yr}$, and for primary CE. The modes are shown with their respective 68\%-range. We caution that most of the PDFs are notably non-normal, often with two dominant peaks at the range limits.
\label{tab:temp_tab_CBP_fast}}
\begin{tabular}{cc||ccc|ccc||c}
\hline
\hline
$\alpha_{CE}$ & Tides &  & ${\rm a_{CBP, PCE}~[AU]}$ & & & ${\rm e_{CBP, PCE}}$ & & Notes \\
 & & ${\rm Mode}$ & ${\rm Median}$ & Range & ${\rm Mode}$ & ${\rm Median}$ & Range &  \\
\hline
\hline
Kepler-38 & & & & & & & & ${\rm [a_{CBP,0} = 0.46]}$ \\
0.5 & NTCP & $0.6^{+0.2}_{-0.01}$& 0.8 & 0.58--1.7 & $0.15^{+0.12}_{-0.07}$ & 0.26 & 0.05--0.67 & \\
0.5$^\dagger$ & TCP & $0.5^{+0.4}_{-0.01}$ & 0.6 & 0.49--0.9 & $0.4^{+0.01}_{-0.35}$ & 0.2 & 0.03--0.42 & \\
1/3/5/10 & NTCP & $1.3^{+0.1}_{-0.2}$ & 1.3 & 0.9--3.8 & $0.38^{+0.16}_{-0.01}$ & 0.42 & 0.02--0.82 & \\
1/3/5/10 & TCP & $1.4^{+0.01}_{-0.2}$ & 1.4 & 1.1--1.7& $0.46^{+0.01}_{-0.08}$ & 0.42 & 0.28--0.56 & \\
\hline
Kepler-47 & & & & & & & & \\
0.5, P1$^{\S}$ & NTCP & $0.31^{+0.09}_{-0.01}$& 0.34 & 0.3--0.44 & $0.12^{+0.04}_{-0.03}$ & 0.14 & 0.06--0.27 & ${\rm [a_{CBP,0} = 0.29]}$\\
0.5, P2$^{\S}$ & NTCP & $0.66^{+0.42}_{-0.01}$& 0.88 & 0.61--1.67 & $0.14^{+0.35}_{-0.05}$ & 0.22 & 0.06--0.56 & ${\rm [a_{CBP,0} = 0.70]}$\\
0.5, P3$^{\S}$ & NTCP & $0.9^{+0.61}_{-0.01}$& 1.2 & 0.8--2.87 & $0.17^{+0.29}_{-0.05}$& 0.23 & 0.09--0.66 & ${\rm [a_{CBP,0} = 0.96]}$\\
0.5$^\dagger$, P1$^{\S}$ & TCP & $0.32^{+0.07}_{-0.01}$ & 0.34 & 0.31--0.39 & $0.1^{+0.08}_{-0.02}$ & 0.1 & 0.07--0.18 & \\
0.5$^\dagger$, P2$^{\S}$ & TCP& $0.7^{+0.5}_{-0.01}$& 0.87 & 0.67--1.24 & $0.08^{+0.32}_{-0.01}$ & 0.21 & 0.06--0.42 & \\
0.5$^\dagger$, P3$^{\S}$ & TCP & $0.93^{+0.91}_{-0.01}$& 1.18 & 0.88--1.89 & $0.11^{+0.35}_{-0.01}$ & 0.22 & 0.09--0.48 & \\
1.0, P1$^{\S}$ & NTCP & $ 0.49^{+0.01}_{-0.02} $& 0.49 & 0.47--0.5 & $0.08^{+0.05}_{-0.04}$& 0.08 & 0.04--0.13 & \\
1.0, P2$^{\S}$ & NTCP & $1.33^{+0.17}_{-0.17}$& 1.41 & 1.12--1.97 & $0.42^{+0.11}_{-0.14}$& 0.4 & 0.22--0.58 & \\
1.0, P3$^{\S}$ & NTCP & $1.65^{+0.53}_{-0.01}$& 2.13 & 1.52--4.15 & $0.5^{+0.18}_{-0.14}$& 0.51 & 0.29--0.75 & \\
1.0$^\dagger$, P1 & TCP & $0.38^{+0.03}_{-0.01}$ & 0.39 & 0.37---0.41 & $0.11^{+0.01}_{-0.1}$ & 0.07 & 0.01--0.12 & \\
1.0$^\dagger$, P2 & TCP & $0.91^{+0.29}_{-0.01}$ & 1.03 & 0.89--1.22 & $0.36^{+0.01}_{-0.21}$ & 0.26 & 0.14--0.38 & \\
1.0$^\dagger$, P3 & TCP & $1.23^{+0.47}_{-0.01}$ & 1.46 & 1.2--1.83 & $0.44^{+0.01}_{-0.26}$ & 0.32 & 0.17--0.45 & \\
3/5/10, P1$^{\S\S}$ & NTCP & $0.67^{+0.01}_{-0.01}$ & 0.67 & 0.66--0.68 & $0.11^{+0.03}_{-0.01}$ & 0.11 & 0.04--0.17 & \\
3/5/10, P2$^{\S\S}$ & NTCP & $3.23^{+0.57}_{-0.01}$ & 3.4 & 2.23--7.95 & $0.71^{+0.1}_{-0.06}$ & 0.73 & 0.57--0.89 & \\
3/5/10, P3$^{\S\S}$ & NTCP & $7.64^{+7.21}_{-2.71}$ & 9.97 & 4.48--22.52 & $0.92^{+0.02}_{-0.02}$ & 0.89 & 0.74--0.95$^\ddagger$ & 46\% ejected\\
3/5/10, P1$^{\S\S}$ & TCP & $0.69^{+0.01}_{-0.01}$ & 0.69 & 0.68--0.7& $0.1^{+0.03}_{-0.01}$ & 0.11 & 0.08--0.14 & \\
3/5/10, P2$^{\S\S}$ & TCP & $3.68^{+0.91}_{-0.23}$ & 4.18 & 3.34--5.62& $0.75^{+0.07}_{-0.01}$ & 0.78 & 0.72--0.83 & \\
3/5/10, P3$^{\S\S}$ & TCP & $15.69^{+6.16}_{-0.01}$ & 19.7 & 15.3--23.01& $0.93^{+0.02}_{-0.01}$ & 0.94 & 0.93--0.95$^\ddagger$ & 81\% ejected \\
\hline
\hline
\multicolumn{9}{l}{$^\dagger$: Sinusoidal distributions in phase vs final acbp/ecbp (see e.g. Fig \ref{fig:K38prim_merger_a_vs_phase}), typically with double-peaked PDFs.} \\
\multicolumn{9}{l}{$^{\S}$: P1/P2/P3 have 40\%/15\%/75\% probability to become dynamically unstable within 1 Myr after the CE phase.} \\
\multicolumn{9}{l}{$^\ddagger$: Given the large parameter space and the associated uncertainties, we consider $\ep>0.95$ as ejection.} \\
\multicolumn{9}{l}{$^{\S\S}$: P1/P2/P3 have 60\%/15\%/85\% probability to become dynamically unstable within 1 Myr after the CE phase.} \\
\end{tabular}
\\
\end{center}
\end{table}}

\clearpage
{\renewcommand{\arraystretch}{1.55}%
\begin{table}[ht]
\begin{center}
\scriptsize
\caption{Same as Table \ref{tab:temp_tab_CBP_fast}.
\label{tab:temp_tab_CBP_fast_cont}}
\begin{tabular}{cc||ccc|ccc||c}
\hline
\hline
$\alpha_{CE}$ & Tides &  & ${\rm \aPCE~[AU]}$ & & & ${\rm \ePCE}$ & & Notes \\
 & & ${\rm Mode}$ & ${\rm Median}$ & Range & ${\rm Mode}$ & ${\rm Median}$ & Range &  \\
\hline
\hline
Kepler-64 & & & & & & & & ${\rm [a_{CBP,0} = 0.65]}$ \\
0.5 & NTCP & $0.8^{+0.6}_{-0.2}$ & 0.9 & 0.6--4.8& $0.16^{+0.24}_{-0.}$ & 0.35 & 0.03--0.85 & \\
0.5$^\dagger$ & TCP & $0.6^{+0.6}_{-0.0}$ & 0.8 & 0.6--1.2& $0.19^{+0.23}_{-0.07}$ & 0.19 & 0.11--0.43 &  \\
1.0 & NTCP & $1.4^{+1.1}_{-0.0}$ & 2.0 & 1.3--7.5 & $0.62^{+0.15}_{-0.31}$ & 0.53 & 0.12--0.8 & \\
1.0$^\dagger$ & TCP & $0.9^{+0.1}_{-0.0}$ & 0.9 & 0.9--1.0& $0.2^{+0.0}_{-0.18}$ & 0.13 & 0.01--0.21 &  \\
3/5/10 & NTCP & $1.6^{+2.6}_{-0.1}$ & 3.1 & 1.4--19.6&  $0.8^{+0.11}_{-0.26}$ & 0.66 & 0.04--0.95$^\ddagger$ & 4\% ejected \\
3/5/10 & TCP & $2.4^{+0.}_{-0.1}$ & 2.4 & 2.2--2.6 & $0.49^{+0.01}_{-0.09}$ & 0.44 & 0.38--0.5 & \\
\hline
Kepler-1647$^{\S}$ & & & & & & & & ${\rm [a_{CBP,0} = 2.71]}$ \\
0.5 & NTCP & $1.6^{+2.3}_{-0.0}$ & 2.4 & 1.4--39.5 & $0.72^{+0.06}_{-0.43}$ & 0.69 & 0.3--0.95 & 47\% ejected\\
0.5$^\dagger$ & TCP & $2.3^{+10.6}_{-0.0}$ & 3.6 & 2.0--13.2 & $0.26^{+0.5}_{-0.01}$ & 0.33 & 0.23--0.79 &  \\
1.0 & NTCP & $1.9^{+2.1}_{-0.0}$ & 2.6 & 1.7--37.5& $0.53^{+0.11}_{-0.11}$ & 0.57 & 0.4--0.95 & 47\% ejected\\
1.0$^\dagger$ & TCP & $3.9^{+5.6}_{-0.0}$ & 5.5 & 3.8--9.7& $0.29^{+0.4}_{-0.0}$ & 0.5 & 0.27--0.71 &   \\
3/5/10 & NTCP & $2.3^{+1.5}_{-0.3}$ & 3.1 & 1.9--47.1& $0.4^{+0.12}_{-0.12}$ & 0.44 & 0.16--0.95 & 54\% ejected\\
3/5/10 & TCP & $9.6^{+4.6}_{-1.6}$ & 11.5 & 7.6--23.2& $0.7^{+0.14}_{-0.01}$ & 0.76 & 0.63--0.88 & \\
\hline
Kepler-1647$^{\S\S}$ & & & & & & & & \\
5.0$^\dagger$  & NTCP & $11.3^{+5.3}_{-1.1}$ & 15.3 & 10.2-53.3 & $0.76^{+0.03}_{-0.03}$& 0.79 & 0.68--0.95 & 86\% ejected\\
10.0$^\dagger$ & NTCP & $8.2^{+6.4}_{-0.6}$ & 10.5 & 7.6--36.9 & $0.64^{+0.07}_{-0.07}$ & 0.71 & 0.57--0.95 & 83\% ejected  \\
10.0 & TCP & $14.5^{+11.1}_{-1.0}$ & 18.2 & 13.5--96.7 & $0.52^{+0.28}_{-0.4}$ & 0.5 & 0.01--0.95 & 85\% ejected\\
\hline
Kepler-35$^{\star}$ & & & & & & & & ${\rm [a_{CBP,0} = 0.60]}$ \\
0.5 & NTCP & $0.87^{+0.23}_{-0.01}$& 1.04 & 0.75--10.13 & $0.44^{+0.27}_{-0.18}$ & 0.41 & 0.03--0.93 & 4\% ejected\\
0.5$^\dagger$ & TCP & $0.87^{+0.47}_{-0.01}$ & 1.01 & 0.84--1.37 & $0.43^{+0.01}_{-0.37}$ & 0.25 & 0.04--0.45 & \\
1/3/5/10 & NTCP & $0.84^{+0.16}_{-0.01}$ & 1.03 & 0.77--6.93 & $0.43^{+0.01}_{-0.26}$ & 0.36 & 0.04--0.91 & \\
1/3/5/10 & TCP & $1.0^{+0.03}_{-0.04}$ & 0.98 & 0.86--1.24& $0.18^{+0.04}_{-0.04}$ & 0.19 & 0.01--0.39 & \\
\hline
\hline
\multicolumn{9}{l}{$^\dagger$: Sinusoidal-like distributions in  \aPCE/\ePCE as a function of $\Delta\theta_0$; typically with double-peaked PDFs.} \\
\multicolumn{9}{l}{$^\ddagger$: Given the large parameter space and the associated uncertainties, we consider $e>0.95$ as ejection.} \\
\multicolumn{9}{l}{$^{\S}$: Primary RLOF and CE.} \\
\multicolumn{9}{l}{$^{\S\S}$: Secondary RLOF and CE.} \\
\multicolumn{9}{l}{$^{\star}$: Experiences CE for ${\rm Z - 1\sigma_Z}$.} \\
\end{tabular}
\\
\end{center}
\end{table}}

\clearpage
{\renewcommand{\arraystretch}{1.55}%
\begin{table}[ht]
\begin{center}
\scriptsize
\caption{Same as Table \ref{tab:temp_tab_CBP_fast} but for ${\rm \aM = 0.1~M_{Sun}/yr}$. Kepler-38b, -47bcd, and -64b evolve adiabatically. Unless indicated otherwise, the spread of their semi-major axis and eccentricity range is practically zero, and for simplicity omitted.
\label{tab:temp_tab_CBP_slow}}
\begin{tabular}{cc||ccc|ccc||c}
\hline
\hline
$\alpha_{CE}$ & Tides &  & ${\rm \aPCE~[AU]}$ & & & ${\rm \ePCE}$ & & Notes \\
 & & ${\rm Mode}$ & ${\rm Median}$ & Range & ${\rm Mode}$ & ${\rm Median}$ & Range &  \\
\hline
\hline
Kepler-38 & & & & & & & & ${\rm [a_{CBP,0} = 0.46]}$\\
0.5 & NTCP & $0.73$& 0.73 & & $0.12$ & 0.12 & & \\
0.5$^\dagger$ & TCP & $0.6$ & 0.61 & & $0.03$ & 0.03 & & \\
1/3/5/10 & NTCP & $1.06$ & 1.06 & & $0.13$ & 0.13 & & \\
1/3/5/10 & TCP & $1.1$ & 1.11 & & $0.02$ & 0.02 & & \\
\hline
Kepler-47 & & & & & & & & \\
0.5, P1 & NTCP & $0.34$& 0.34 & & $0.04$ & 0.04 & & ${\rm [a_{CBP,0} = 0.29]}$\\
0.5, P2 & NTCP & $0.83$& 0.83 & & $0.04$ & 0.04 &  &${\rm [a_{CBP,0} = 0.70]}$ \\
0.5, P3 & NTCP & $1.13$& 1.14 & & $0.06$ & 0.06 & & ${\rm [a_{CBP,0} = 0.96]}$\\
0.5$^\dagger$, P1 & TCP & $0.34$ & 0.34 & & $0.01$ & 0.01 &  & \\
0.5$^\dagger$, P2 & TCP & $0.8$& 0.83 & & $0.02$& 0.02 &  & \\
0.5$^\dagger$, P3 & TCP & $1.13$& 1.14 & & $0.03$& 0.03 & & \\
1.0, P1 & NTCP & $ 0.48 $& 0.48 & & $0.03$ & 0.03 &  & \\
1.0, P2 & NTCP & $1.18$& 1.18 & & $0.05$ & 0.05 &  & \\
1.0, P3 & NTCP & $1.6$& 1.61 & & $0.06$ & 0.06 &  & \\
1.0$^\dagger$, P1 & TCP & $0.39$ & 0.39 & & $0.01$ & 0.01 & & \\
1.0$^\dagger$, P2 & TCP & $0.96$ & 0.96 & & $0.02$ & 0.02 &  & \\
1.0$^\dagger$, P3 & TCP & $1.3$ & 1.31 & & $0.03$ & 0.03 &  & \\
3/5/10, P1 & NTCP & $0.66$ & 0.66 & & $0.04$ & 0.04 & &\\
3/5/10, P2 & NTCP & $1.61$ & 1.61 & & $0.07$ & 0.07 &  & \\
3/5/10, P3 & NTCP & $2.2$ & 2.2 & & $0.08$ & 0.08 & &\\
3/5/10, P1 & TCP & $0.69$ & 0.69 & & $0.01$ & 0.01 &  &  \\
3/5/10, P2 & TCP & $1.67$ & 1.67 & & $0.03$ & 0.03 & &  \\
3/5/10, P3 & TCP & $2.28$ & 2.28 & & $0.04$ & 0.04 & &  \\
\hline
\hline
 \multicolumn{9}{l}{$^\dagger$: Sinusoidal distributions in phase vs final acbp/ecbp (see Fig \ref{fig:K38prim_merger_a_vs_phase}), typically with double-peaked PDFs.} \\
 \multicolumn{9}{l}{$^\ddagger$: Given the large parameter space and the associated uncertainties, we consider $e>0.95$ as ejection.} \\
\end{tabular}
\\
\end{center}
\end{table}}

\clearpage
{\renewcommand{\arraystretch}{1.55}%
\begin{table}[ht]
\begin{center}
\scriptsize
\caption{Same as Table \ref{tab:temp_tab_CBP_slow}. 
\label{tab:temp_tab_CBP_slow_cont}}
\begin{tabular}{cc||ccc|ccc||c}
\hline
\hline
$\alpha_{CE}$ & Tides &  & ${\rm \aPCE~[AU]}$ & & & ${\rm \ePCE}$ & & Notes \\
 & & ${\rm Mode}$ & ${\rm Median}$ & Range & ${\rm Mode}$ & ${\rm Median}$ & Range &  \\
\hline
\hline
Kepler-64 & & & & & & & & ${\rm [a_{CBP,0} = 0.65]}$\\
0.5 & NTCP & $0.82$ & 0.82 & & $0.25$ & 0.25 & 0.22--0.27 & \\
0.5$^\dagger$ & TCP & $0.72$ & 0.72 & & $0.02$ & 0.02 &  &  \\
1.0 & NTCP & $1.58$ & 1.58 & & $0.31$ & 0.33 & 0.31--0.35 & \\
1.0$^\dagger$ & TCP & $0.91$ & 0.91 && $0.01$ & 0.01 & &  \\
3/5/10 & NTCP & $1.83$ & 1.83 & &  $0.3$ & 0.35 & 0.31--0.39 & \\
3/5/10 & TCP & $1.9$ & 1.9 & & $0.02$ & 0.02 &  & \\
\hline
Kepler-1647$^{\S}$ & & & & & & & & ${\rm [a_{CBP,0} = 2.71]}$\\
0.5 & NTCP & $2.52^{+1.53}_{-0.01}$ & 3.55 & 2.33--40.61 & $0.7^{+0.02}_{-0.08}$ & 0.71 & 0.62--0.95$^\ddagger$ & 27\% ejected\\
0.5$^\dagger$ & TCP & $3.1^{+0.19}_{-0.01}$ & 3.19 & 3.08--3.3 & $0.11^{+0.01}_{-0.08}$ & 0.07 & 0.02--0.12 &  \\
1.0 & NTCP & $3.91^{+0.67}_{-0.01}$ & 4.45 & 3.83--20.27 & $0.72^{+0.22}_{-0.01}$ & 0.81 & 0.71--0.95 & 27\% ejected\\
1.0$^\dagger$ & TCP & $4.1$ & 4.1 & & $0.07$ & 0.07 &  &   \\
3/5/10 & NTCP & $4.7^{+0.63}_{-0.26}$ & 4.93 & 4.29--19.14 & $0.7^{+0.19}_{-0.01}$ & 0.76 & 0.57--0.95 & 4\% ejected\\
3/5/10 & TCP & $4.93$ & 4.93 & & $0.08$ & 0.08 & & \\
\hline
Kepler-1647$^{\S\S}$ & & & & & & & & \\
5.0$^{\dagger\dagger}$  & NTCP & $7.9^{+6.3}_{-0.7}$ & 9.9 & 7.2--55.9 & $0.48^{+0.9}_{-0.1}$ & 0.55 & 0.2--0.95 & 82\% ejected\\
10.0$^{\dagger\dagger}$ & NTCP & $7.7^{+4.0}_{-0.8}$ & 9.2 & 6.9--31.1 & $0.32^{+0.25}_{-0.08}$ & 0.54 & 0.24--0.95 & 81\% ejected  \\
10.0 & TCP & -- & -- & -- & -- & -- & -- & 100\% ejected\\
\hline
Kepler-35$^{\star}$ & & & & & & & & ${\rm [a_{CBP,0} = 0.60]}$ \\
0.5 & NTCP & $0.94$& 0.94 && $0.37$ & 0.37 & 0.33--0.41 & \\
0.5$^\dagger$ & TCP & $0.95$ & 0.95 &  & $0.03$ & 0.03 & & \\
1/3/5/10 & NTCP & $0.94$ & 0.94 &  & $0.27$ & 0.27 & 0.18--0.39 & \\
1/3/5/10 & TCP & $0.95$ & 0.95 &  & $0.02$ & 0.02 & & \\
\hline
\hline
\multicolumn{9}{l}{$^\ddagger$: Given the large parameter space and the associated uncertainties, we consider $e>0.95$ as ejection.} \\
\multicolumn{9}{l}{$^{\S}$: Primary RLOF and CE.} \\
\multicolumn{9}{l}{$^{\S\S}$: Secondary RLOF and CE.} \\
\multicolumn{9}{l}{$^{\star}$: Experiences CE for ${\rm Z - 1\sigma_Z}$.} \\
\multicolumn{9}{l}{$^{\dagger\dagger}$: Non-normal PDF with multiple peaks of comparable strength, mode not well defined.} \\
\end{tabular}
\\
\end{center}
\end{table}}

\begin{table}[ht]
\begin{center}
\caption{Expected dynamical and light travel-time ETV amplitudes and periods after the respective CE stages for {\it Kepler}~CBPs with known masses (i.e. Kepler-1647 and Kepler-35). For comparison, we also list the detected ETV amplitudes and periods for the CBP candidates in NN Ser and HW Vir.
\label{tab:etvs}}
\begin{tabular}{c|ccc|c}
\hline
\hline
${\rm \alpha_{M}}$ & ${\rm A_{dyn} (min/max)}$ & ${\rm A_{LTTE}(min/max)}$ & P(min/max) & Notes \\
$[{\rm \Msun/yr}]$ & [sec] & [sec] & [yr] & \\
\hline
\hline
Kepler-1647b$^\dagger$ & & & & \\
1.0 & 0.2/13.4$^{\ddagger}$ & 68/1682 & 2.4/292 & \\
0.1 & 0.1/13 & 154/680 & 8.1/75 & \\
\hline
Kepler-1647b$^{\dagger\dagger}$ & & & & \\
1.0 & 0./0.2 & 736/9680 & 31/1490 & \\
0.1 & 0./0.2 & 668/5420 & 27/624 & \\
\hline
Kepler-35b$^\dagger$ & & & & \\
1.0 & 0.1/56.5 & 4.3/36.9 & 0.7/17 & \\
0.1 & 0.2/44.2 & 5.0/5.1 & 0.9/0.9 & \\
\hline
NN Ser bc & & & & \\
-- & $\sim5/30$$^{\S}$ & $\sim5/30$ & $\sim7/15$ & \\
\hline
HW Vir bc & & & & \\
-- & $\sim55/550$$^{\S}$ & $\sim50/550$ & $\sim13/55$ & \\
\hline
\hline
\multicolumn{5}{l}{$\dagger$: After primary CE.} \\
\multicolumn{5}{l}{$^{\ddagger}$: Min/max for the range of parameters from Table \ref{tab:temp_tab_CBP_fast_cont}.} \\
\multicolumn{5}{l}{$\dagger\dagger$: After secondary CE.} \\
\multicolumn{5}{l}{${\S}$: Showing the combined effect of inner/outer CBPs (instead of min/max).} \\
\end{tabular}
\\
\end{center}
\end{table}

\end{document}